\documentclass[longauth]{aa} 
\usepackage{graphicx}
\usepackage{natbib} 
\bibpunct{(}{)}{;}{a}{}{,} 

\usepackage{txfonts}
\usepackage{amssymb}
\def\gapprox{\;\rlap{\lower 2.5pt            
 \hbox{$\sim$}}\raise 1.5pt\hbox{$>$}\;}
\def\lapprox{\;\rlap{\lower 2.5pt            
 \hbox{$\sim$}}\raise 1.5pt\hbox{$<$}\;}

\begin{document}

   \title{SPHERE / ZIMPOL observations of the symbiotic system R Aqr}

   \subtitle{I. Imaging of the stellar binary and the innermost jet clouds}

   \author{
        H.M.~Schmid\inst{\ref{instch1}} 
   \and A.~Bazzon\inst{\ref{instch1}} 
   \and J.~Milli\inst{\ref{insteso2}}           
 \and R.~Roelfsema\inst{\ref{instnl1}}   
   \and N.~Engler\inst{\ref{instch1}}
   \and D.~Mouillet\inst{\ref{instf1},\ref{instf2}} 
   \and E.~Lagadec\inst{\ref{instf5}}      
   \and E.~Sissa\inst{\ref{insti1},\ref{insti2}}
   \and J.-F.~Sauvage\inst{\ref{instf3}}        
   \and C.~Ginski\inst{\ref{instnl3},\ref{instnl2}}      
   \and A.~Baruffolo\inst{\ref{insti1}}
   \and J.L.~Beuzit\inst{\ref{instf1},\ref{instf2}}
   \and A.~Boccaletti\inst{\ref{instf4}}      
   \and A.J.~Bohn\inst{\ref{instch1}}
   \and R.~Claudi\inst{\ref{insti1}} 
   \and A.~Costille\inst{\ref{instf3}}     
   \and S.~Desidera\inst{\ref{insti1}}
   \and K.~Dohlen\inst{\ref{instf3}}     
   \and C.~Dominik\inst{\ref{instnl2}}
   \and M.~Feldt\inst{\ref{instd1}}      
   \and T.~Fusco\inst{\ref{instf6}}     
   \and D.~Gisler\inst{\ref{instd2}}      
   \and J.H.~Girard\inst{\ref{insteso2}}      
   \and R.~Gratton\inst{\ref{insti1}}
   \and T.~Henning\inst{\ref{instd1}}      
   \and N.~Hubin\inst{\ref{insteso1}}      
   \and F.~Joos\inst{\ref{instch1}}      
   \and M.~Kasper\inst{\ref{insteso1}}   
   \and M.~Langlois\inst{\ref{instf7},\ref{instf3}}      
   \and A.~Pavlov\inst{\ref{instd1}}
   \and J.~Pragt\inst{\ref{instnl1}}       
   \and P.~Puget\inst{\ref{instf1}} 
   \and S.P.~Quanz\inst{\ref{instch1}}            
   \and B.~Salasnich\inst{\ref{insti1}}
   \and R.~Siebenmorgen\inst{\ref{insteso1}}
   \and M.~Stute\inst{\ref{instd3}}
   \and M.~Suarez\inst{\ref{insteso1}}
   \and J.~Szul\'{a}gyi\inst{\ref{instch1}} 
   \and C.~Thalmann\inst{\ref{instch1}}  
   \and M.~Turatto\inst{\ref{insti1}}   
   \and S.~Udry\inst{\ref{instch2}} 
   \and A.~Vigan\inst{\ref{instf3}}        
   \and F.~Wildi\inst{\ref{instch2}} 
          }

\institute{
ETH Zurich, Institute for Astronomy, Wolfgang-Pauli-Strasse 27, 
CH-8093 Zurich, Switzerland\label{instch1}
\and
European Southern Observatory, Alonso de Cordova 3107, Casilla
19001 Vitacura, Santiago 19, Chile\label{insteso2}
\and
NOVA Optical Infrared Instrumentation Group at ASTRON, Oude
Hoogeveensedijk 4, 7991 PD Dwingeloo, The Netherlands\label{instnl1}
\and
Universit\'{e} Grenoble Alpes, IPAG, 38000 Grenoble, France\label{instf1}
\and
CNRS, IPAG, 38000 Grenoble, France\label{instf2}
\and
Universit\'{e} C\^{o}te d'Azur, Observatoire de la C\^{o}te d'Azur, CNRS, 
Lagrange, France\label{instf5}
\and
INAF – Osservatorio Astronomico di Padova, Vicolo
dell’Osservatorio 5, 35122 Padova, Italy\label{insti1}
\and
Dipartimento die Fisica e Astronomia ``G. Galilei'', 
Universit\'{a} di Padova, Vicolo dell’Osservatorio 5, 35122 Padova, 
Italy\label{insti2}
\and
Aix Marseille Univ, CNRS, LAM, Laboratoire d'Astrophysique de Marseille, Marseille, France\label{instf3}
\and
Leiden Observatory, Leiden University, P.O. Box 9513, 2300 RA
Leiden, The Netherlands\label{instnl3}
\and
Anton Pannekoek Astronomical Institute, University of Amsterdam,
PO Box 94249, 1090 GE Amsterdam, The Netherlands\label{instnl2}
\and
LESIA, Observatoire de Paris, PSL Research University, 
CNRS, Sorbonne Universités, UPMC Univ. Paris 06, Univ. Paris Diderot, 
Sorbonne Paris Cité, 5 place Jules Janssen, 92195 Meudon, France\label{instf4}
\and
Max-Planck-Institut f\"{u}r Astronomie, K\"{o}nigstuhl 17, 69117
Heidelberg, Germany\label{instd1}
\and
ONERA, The French Aerospace Lab BP72, 29 avenue de la
Division Leclerc, 92322 Ch\^{a}tillon Cedex, France\label{instf6}
\and
Kiepenheuer-Institut f\"{u}r Sonnenphysik, Schneckstr. 6, D-79104
Freiburg, Germany\label{instd2}
\and
European Southern Observatory, Karl Schwarzschild St, 2, 85748
Garching, Germany\label{insteso1}
\and
Centre de Recherche Astrophysique de Lyon, CNRS/ENSL
Universit\'{e} Lyon 1, 9 av. Ch. Andr\'{e}, 69561 Saint-Genis-Laval,
France\label{instf7}
\and
Simcorp GmbH, Justus-von-Liebig-Strasse 1, D-61352 Bad Homburg, Germany\label{instd3}
\and
Geneva Observatory, University of Geneva, Chemin des Mailettes
51, 1290 Versoix, Switzerland\label{instch2}
            }

   \date{Received ...; accepted ...}

\abstract
{R Aqr is a symbiotic binary system consisting of a mira variable, 
a hot companion with a spectacular jet outflow, and an extended 
emission line nebula. Because of its proximity to the sun, this object has 
been studied in much detail with many types of high resolution imaging and 
interferometric techniques. We have used 
R Aqr as test target for the visual camera subsystem ZIMPOL, which is 
part of the new extreme adaptive optics (AO) instrument SPHERE at the 
Very Large Telescope (VLT).} 
{We describe SPHERE-ZIMPOL test observations of the
R Aqr system taken in H$\alpha$ and other filters in order to 
demonstrate the exceptional performance of this high resolution 
instrument. We compare our observations with data from the 
Hubble Space Telescope (HST) and
illustrate the complementarity of the two instruments. We use our
data for a detailed characterization of the inner jet region of R Aqr. 
}
%
{We analyze the high resolution $\approx 25$ mas images from  
SPHERE-ZIMPOL and determine from the H$\alpha$ emission
the position, size, geometric structure, and 
line fluxes of the jet source and the clouds in the
innermost region $<2''$ ($<400$~AU) of R Aqr.
The data are compared to simultaneous HST line filter observations. 
The H$\alpha$ fluxes and the measured sizes of the clouds yield  
H$\alpha$ emissivities for many clouds from which one can
derive the mean density, mass, recombination time scale, 
and other cloud parameters.}
%
{Our H$\alpha$ data resolve for the first time the R Aqr binary
and we measure for the jet source a relative position $46\pm 1$ 
mas West (position angle $-85.5^\circ\pm 1.0^\circ$) of the mira. The central jet source is the strongest
H$\alpha$ component with a flux
of about $2.5\cdot 10^{-12}\,{\rm erg}\,{\rm cm}^{-2}{\rm s}^{-1}$. 
North east and south west from the central source there are many
clouds with very diverse structures. Within 
$0.5''$ (100 AU) we see in the SW a 
string of bright clouds arranged in a zig-zag pattern 
and, further out, at $1'' - 2''$, fainter and more extended bubbles. 
In the N and NE we see a
bright, very elongated filamentary structure between $0.2''-0.7''$ 
and faint perpendicular ``wisps'' further out. \\
Some jet clouds are also detected in the ZIMPOL [O~I]  
and He~I filters, as well as in the HST-WFC3 line filters for H$\alpha$,
[O~III], [N~II], and [O~I]. We determine jet cloud parameters 
and find a very well defined correlation $N_e\propto r^{-1.3}$ between
cloud density and distance to the central binary. Densities are 
very high with typical values of $N_e\approx 3\cdot 10^5{\rm cm}^{-3}$ for the
``outer'' clouds around 300~AU,  
$N_e\approx 3\cdot 10^6{\rm cm}^{-3}$ for the ``inner'' clouds 
around 50~AU, and even higher for the central jet source.
The high $N_e$ of the clouds implies short recombination or variability
timescales of a year or shorter.} 
%
{H$\alpha$ high resolution data provide a lot of diagnostic information 
for the ionized jet gas in R Aqr. Future H$\alpha$ observations will 
provide the orientation of the orbital plane of the binary and 
allow detailed hydrodynamical investigations of this jet outflow and
its interaction with the wind of the red giant companion.}


\authorrunning{H.M. Schmid et al.}

\titlerunning{SPHERE / ZIMPOL observations of the symbiotic system R Aqr}

\maketitle
%

\begin{figure*}
\includegraphics[trim=0.5cm 0.5cm 0.5cm 0.5cm,clip,width=18cm]{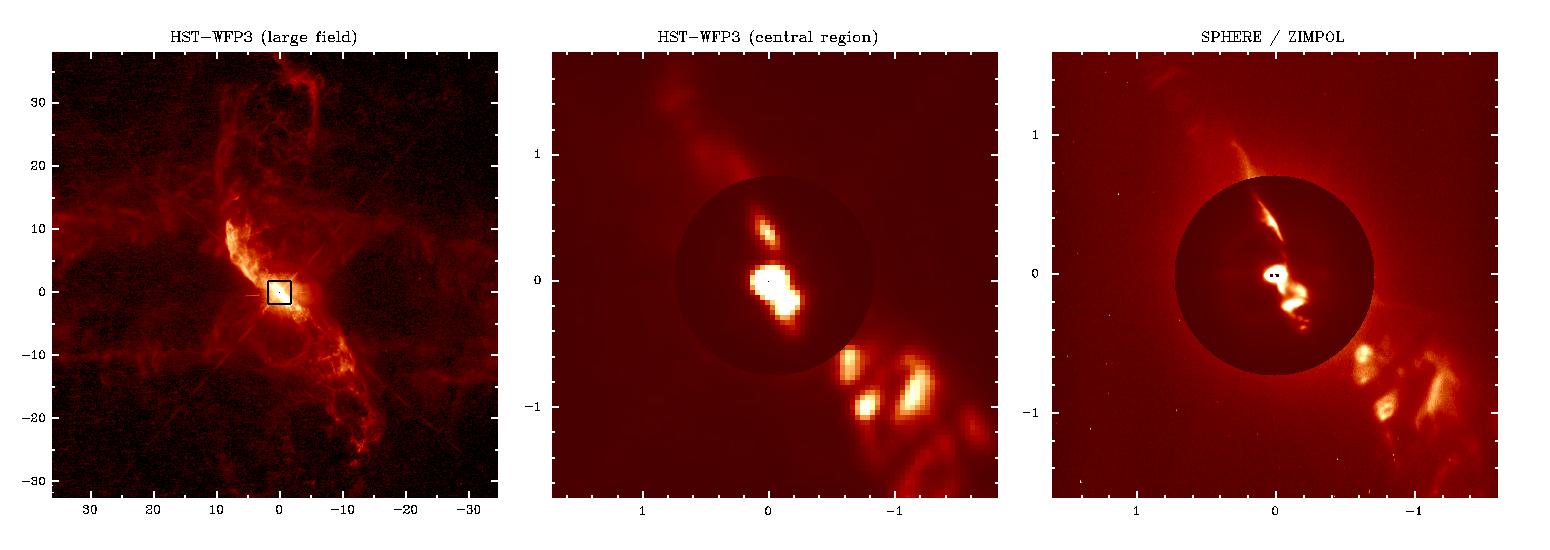}
\caption{H$\alpha$ images of R Aqr from HST-WFC3 and VLT-SPHERE-ZIMPOL taken 
in October 2014. The left panel shows a $70'' \times 70''$ cutout of the
WFC3 image of the strongly structured extended nebula. The middle 
panel is the $3.5''\times 3.5''$ region 
of the WFC3 image of the central star indicated with the square in the 
left panel. The right panel is the central $3.2''\times 3.2''$ 
area imaged with higher spatial resolution using SPHERE-ZIMPOL where
also the two stars are marked with black dots.
The color scale is 10 times enhanced for the central $r<0.7''$ region in
the middle and the right panel. North is up and East to the left.}
\label{HSTZimpol}
\end{figure*}
%

  
\section{Introduction} \label{Introduction}

\object{R Aqr} is a peculiar mira variable with a pulsation
period of 387 days surrounded by an extended emission line nebulosity 
\citep[e.g.,][]{Lampland22,Hollis99}.
Detailed studies in many wavelength bands revealed that
R Aqr is a symbiotic binary with a mass-losing, pulsating red giant 
and an accreting hot companion with a jet outflow which ionizes an
emission nebula. R Aqr is thus an interesting system,
and because of its proximity to the sun it became a prototype
object for studies on stellar jet, symbiotic (nova-like)
activity, mass transfer, and mass loss in interacting 
binaries. 

The orbital period of the R Aqr binary is about $P\approx 44$~years 
as inferred from periodic phases of reduced brightness observed 
around 1890, 1933, and 1977 \citep{Willson81}. These phases are 
interpreted as partial obscurations of the mira by the  
companion with its accretion disk and the associated gas and dust flows.
The inferred orbital period is supported by radial velocity measurements
\citep[see][]{Hinkle89,Gromadzki09a}.

The R Aqr system was extensively studied with many kinds of high resolution
imaging techniques since the first reports about the appearance of 
a ``brilliant emission jet or spike'' in 1980 \citep{Wallerstein80,Herbig80}. 
The structure and motion of jet outflow features was observed with
long slit spectroscopy \citep{Solf85}, radio interferometry 
\citep[e.g.,][]{Hollis85,Kafatos89,Dougherty95},
imaging with the Hubble Space Telescope (HST) \citep{Paresce94,Hollis97a}, 
and Chandra X-ray observations 
\citep{Kellogg01}, while the photosphere of the mira variable 
and its immediate surroundings were
investigated with maser line radio interferometry 
\citep[e.g.,][]{Hollis01,Cotton04,Ragland08,Kamohara10}, and infrared
(IR) interferometry \citep{Ragland08,Zhao-Geisler12}. 

We present new high resolution observations of the central
jet outflow of R Aqr taken with the new SPHERE
(the {\sl S}pectro-{\sl P}olarimetric {\sl H}igh-contrast 
{\sl E}xoplanet {\sl RE}search) ``Planet Finder'' instrument at 
the Very Large Telescope (VLT) \citep{Beuzit08}. 
The {\sl Z}urich {\sl IM}aging {\sl POL}arimeter (ZIMPOL), 
the visible camera subsystem 
of SPHERE, provides imaging (and polarimetric imaging) with a resolution 
of about 25 mas for nebular lines, in particular the prominent 
H$\alpha$ emission. R Aqr was observed during the instrument commissioning, 
because this bright star with circumstellar emission is ideal 
for on-sky tests of line filter observations and imaging polarimetry. 
Fortunately, one of our test runs
took place just a few days before HST - WFC3
line filter observation or R Aqr. This provides
a unique opportunity for improving the ZIMPOL flux
measurements and the instrument throughput calibration.  

For the scientific investigation of the R Aqr system the HST and 
SPHERE data are very complementary.  
HST provides a much larger field of view, higher sensitivity, and
flux fidelity, while ZIMPOL-SPHERE yields imaging and 
polarimetric imaging with about three times higher
spatial resolution and higher contrast in a small field
($3.6''\times 3.6''$) centered on the star. The enhanced
resolution enables ZIMPOL-SPHERE to resolve the central binary system,
the innermost jet clouds, and polarized
light produced by the scattering from circumstellar dust particles. 

In this work we concentrate the scientific investigation 
on the SPHERE and HST line filter observations for the 
small central field of R Aqr. Key topics are the imaging of
the central binary and the properties of the innermost
jet clouds. We put particular
emphasis on accurate absolute flux measurements for the central 
jet source and the H$\alpha$ cloud components 
seen in the ZIMPOL data. This is a notoriously 
difficult task for observations taken with ground based adaptive 
optics (AO) systems
and therefore we want to take advantage of the quasi-simultaneous HST data. 
The SPHERE-ZIMPOL imaging polarimetry of R Aqr will be presented 
in a future paper.   

Figure \ref{HSTZimpol} gives a first overview of the new H$\alpha$-maps. 
The HST images from October~2014 show the central part of the 
extended nebulosity consisting of an inclined ring 
with a semi-major axis of about 42 arcsec oriented in an E-W direction 
with an inclination angle of about 70 degrees. 
In Fig.~\ref{HSTZimpol} (left) only the near and far side of the 
apparent ellipse are visible, about 10$''$ above and below the star.
There is also the elongated two-sided jet structure in NE and SW directions and
associated arcs extending to about 30 arcsec from the central source.
The bright ``jet spike'' initially detected around 1980 has moved 
radially out with an angular speed of about $0.2''$/yr 
\citep[e.g.,][]{Maerkinen04} and it is
now the prominent elongated feature at about 10 arcsec to the NE 
of the central source. 

The jet outflow pattern in R Aqr is quite complex with measured
proper motions corresponding to tangential speeds between 50 and 
250~kms/s and radial velocities from $-100$~km/s to $+100$~km/s
\citep{Solf85,Hollis97a,Navarro03,Maerkinen04}. The overall flow 
pattern of the NE jet spike corresponds to an outflow away from the
central binary, with a velocity of about 150~km/s and a radial velocity
component of $-70$~km/s (towards us). The SW jet moves roughly 
in the opposite direction. 

Modeling \citep[e.g.,][]{Burgarella92,Contini03} of mainly the
bright NE-jet feature, but also other clouds at separations of $>1$~arcsec, 
indicates that the ionized gas is produced by shocks 
caused by the interaction of a fast ($v>100$~km/s), collimated outflow 
from the central binary with slower material ($v<40$ km/s) in the system. 

The H$\alpha$ image for the very bright central region 
with the stellar R Aqr source is 
shown in Fig.~\ref{HSTZimpol}b for HST-WFC3 and in Fig.~\ref{HSTZimpol}c 
for SPHERE-ZIMPOL. Because of the higher resolution
of ZIMPOL, it is possible to resolve the central binary (marked
with two dots) and the innermost jet clouds, and we can measure 
for the first time the exact separation and orientation of the
stars.

Jet outflow components at separations of $<0.5''$ from the
mira have been detected previously with HST imaging and 
radio interferometry \citep{Paresce94,Dougherty95}. These studies show
a strong variability of the innermost jet structures. Possibly,
the binary was already previously 
resolved by \citet{Hollis97b} with a map of 
quasi-simultaneous observation of SiO maser emission from the red giant 
located about 50 mas south of an extended radio continuum 
emission, which was associated with the jet source. Unfortunately no second
epoch data were published which confirm this. 
It was also possible to observe the mira photosphere
and circumstellar maser emission with a resolution in the  
1-10 milli-arcsec range
\citep[e.g.,][]{Ragland08}, but the relative location of the companion 
star could not be constrained from such observations. The new data
presented in this work provide images of the central jet outflow with
much improved resolution and sensitivity, and they resolve clearly
the two stellar components in the system.  
 
This paper is organized as follows. Section 2 gives an overview on the 
VLT-SPHERE observations, a description of the used filters, 
and an assessment of the SPHERE AO performance for the R Aqr observations. 
In Section 3 we determine the relative astrometric position for 
the central binary. Section 4 provides the photometry for the mira 
variable in R Aqr and the flux for the total H$\alpha$ emission 
in the ZIMPOL field. The structures of the observed jet clouds
are described in Section 5 including cloud position and size, 
and the derivation of the H$\alpha$ surface brightness and flux for the
individual clouds. Section 6 describes the used HST line filter 
data and the determination of HST line fluxes.
Physical parameters for the jet clouds
are derived and analyzed in Section 7 and the final Section 8 puts our new 
detections on the binary geometry and the jet structure into context 
with previous and future R Aqr observations.

\section{SPHERE / ZIMPOL observations}
\label{SectObs}

\subsection{The SPHERE / ZIMPOL instrument}
\label{SectZimpol}   
The SPHERE ``Planet Finder'' instrument was successfully 
installed and commissioned in 2014 at the VLT. 
SPHERE is optimized for high contrast and high spatial resolution observation
in the near-IR and the visual spectral region using an extreme AO
system, stellar coronagraphs, and three focal plane 
instruments for differential imaging. Technical descriptions
of the instrument are given in, 
for example, \citet{Beuzit08}, \citet{Kasper12}, 
\citet{Dohlen06}, \citet{Fusco14}, and references therein, 
and much basic information can be found in the SPHERE user manual 
and related technical 
websites\footnote{For example, www.eso.org/sci/facilities/paranal.html.} 
of the European Southern Observatory (ESO). 
A first series of SPHERE science papers demonstrate
the performance of various observing modes of this instrument 
\citep[e.g.,][]{Vigan16,Maire16,Zurlo16,Bonnefoy16,Boccaletti15, 
Thalmann15,Kervella16,Garufi16}. 

ZIMPOL is one of 
three focal plane subsystems within SPHERE working in the spectral range from 
520~nm to 900~nm
\citep{Schmid06a,Thalmann08,Roelfsema10,Bazzon12,Schmid12}. 
ZIMPOL provides differential imaging modes 
including angular differential imaging (ADI), 
spectral differential imaging (SDI), and polarimetric 
differential imaging (PDI). It is designed to take advantage of the high spatial resolution ($\approx$ 20-30 mas) offered
by the VLT and the SPHERE extreme AO system, and the high
contrast capabilities of the SPHERE visible coronagraph. 

ZIMPOL has two camera arms, camera 1 and camera 2, and data
are always taken simultaneously in both arms, each 
equipped with its own filter wheel (FW1 and FW2). This allows
us to take data in two different filters simultaneously for
SDI. One can also use two equal filters in the two arms
or use for both detectors the same filter located in wheel FW0 in the 
preceding common path.   

The pixel scale of ZIMPOL
is $3.601\pm 0.005$~mas/pix (mas: milli-arcsec) according to
a preliminary astrometric calibration (Ginski et al., in preparation).
The position angle of the vertical frame axis is $-2.0\pm 0.5$ degrees
with respect to north for both cameras. This offset angle applies 
for preprocessed data\footnote{Preprocessing is the first step in the
data reduction.} which have been flipped up-down and for camera 2 also 
left-right to put N up and E to the left. The field of view of the
1k x 1k detectors is $3.6''\times 3.6''$. Observations are only
possible within $4''$ from a star with an averaged magnitude 
$m\lapprox 10^m$ for the range 500 -- 900 nm, which is bright enough to 
be used as AO wave front source. 

Photometric calibration standard stars are observed regularly
for the throughput calibration of the instrument. In this work we report
photometric zero points for some filters based on preliminary
instrument throughput measurements (Schmid et al., in preparation). Similar high resolution imaging capabilities in
the visual range are currently also offered by 
the VisAO science camera of the MagAO system (without polarimetry)
at the 6.5m Magellan telescope \citep{Close14} and the VAMPIRES
aperture masking interferometer using the SCExAO system at the
Subaru telescope \citep{Norris15}. 

\begin{table*}
\caption{R Aqr observational data from the SPHERE commissioning.}
\label{RAqrdata}
\begin{tabular}{lccccccccc}
\noalign{\smallskip\hrule\smallskip}
frame             & inst/det & \multispan{3}{\hfil filters \hfil} 
                              & DIT & nDIT & nEXP & dc  & remark \\
identifications\tablefootmark{a} 
                  & mode &  FW0 & FW1 & FW2    & [s] &      &   &    & \\
\noalign{\smallskip\hrule\smallskip}

\noalign{\smallskip \noindent 2014-08-12 \smallskip}
OBS224\_0092 & imaging & -- & B\_Ha & Cnt\_Ha 
                   &  100  & 3    & 1   & 0.4 & peak (RG) saturated  \\ 
OBS224\_0093 & imaging & -- & Cnt\_Ha & B\_Ha 
                   &  100  & 3    & 1   & 0.4 & peak (RG) saturated  \\ 
OBS224\_0094 & imaging & -- & Cnt\_Ha & N\_Ha 
                   &  100  & 3    & 1   & 0.4 &  Cnt\_Ha (RG) saturated \\ 
\noalign{\smallskip}
OBS224\_0095 & imaging & V\_S & -- & --        
                      & 20    & 3    & 1   & 0.1 &   \\  
OBS224\_0096 & imaging & OI\_630 & -- & --     
                      & 100   & 3    & 1   & 0.4 &   \\  
OBS224\_0097 & imaging & HeI & -- & --         
                      & 100   & 3    & 1   & 0.4 &   \\  
\noalign{\medskip \noindent 2014-10-11 \smallskip}
OBS284\_0030--34 & imaging & N\_Ha & -- & --       
                   & 40  & 1    & 5    & 0.2 & with dithering \\ 
OBS284\_0035--38 & imaging & N\_Ha & -- & --       
                   & 200 & 1    & 4    & 0.8 & off-axis fields \\ 
OBS284\_0051--54 & slow pol. & -- & CntHa & N\_Ha    
                   & 50  & 2    & 4    & 4.0 &    \\   
\noalign{\smallskip}
OBS284\_0039--42 & fast pol. & -- & V & V            
                      & 1.2 & 10   & 4    & $<0.1$ &          \\  
OBS284\_0055--58 & slow pol. & -- & V & V            
                      & 10  & 4    & 4    & $<0.1$ &         \\  
OBS284\_0043--46 & fast pol. & -- & TiO\_717 & Cnt748 
                      & 1.2 & 10   & 4    &  $<0.1$ &        \\  
OBS284\_0047--50 & fast pol. & -- & Cnt820 & Cnt820  
                      & 1.2 & 10   & 4    &  $<0.1$ &        \\  
\noalign{\smallskip}
OBS284\_0059--62\tablefootmark{b} 
                      & fast pol. & -- & I\_PRIM & I\_PRIM 
                      & 1.2 & 10   & 4    & $<0.1$ & peak saturated \\ 
OBS284\_0063--70\tablefootmark{b} 
                      & fast pol. & -- & I\_PRIM & I\_PRIM 
                      & 5   & 6    & 8    & $<0.1$ & coronagraphic \\ 
OBS284\_0071--74\tablefootmark{b} 
                      & slow pol. & -- & I\_PRIM & I\_PRIM 
                      & 10  & 20   & 4    & $0.8$ & coronagraphic \\ 
\noalign{\smallskip\hrule\smallskip}
\end{tabular}
\tablefoot{The columns list the frame identification, instrument/detector mode, 
the used filters in the filter wheels FW0, FW1, and FW2, detector 
integration times (DIT), number of 
integrations (nDIT) per exposure, the number of exposures (nEXP),
and the estimated dark current (dc) level in ct per pix and frame 
(or per DIT).\\
\tablefoottext{a}{the file identification corresponds to the fits-file
header keyword ``origname'' without prefix ``SPHERE\_ZIMPOL\_''. 
The first three digits give the day of the year followed
by the four-digit observation number.} 
\tablefoottext{b}{data not used in this paper.}
} 
\end{table*}

\subsection{SPHERE / ZIMPOL data}

\object{R Aqr} was used as a test source for the verification
of different instrument configurations and therefore the observations
are not optimized for scientific purposes. The data are 
affected by several technical problems, especially for the July run when
no scientifically useful data could be obtained. In August 2014 only imaging 
observations were possible, but no imaging polarimetry. The data 
from October 2014 could be taken without technical problems.  
For the scientific investigation of R Aqr we use only the 
data from August and October 2014 listed in Table~\ref{RAqrdata}.
All data were taken in field-stabilized mode without 
field position angle offset and using the gray beam splitter
sending about 21~\% of the light to the wave front sensor (WFS) 
and transmitting 79~\% to the ZIMPOL instrument.  

The R Aqr mira variable has a pulsation period of 387.3 days
\citep{Gromadzki09a}. In October 2014 the mira brightness 
was, according to the light curve of the American Association of
Variable Star Observers (AAVSO\footnote{http://www.aavso.org.}),
at its minimum phase with $m_{\rm vis}\approx 11$~mag. This is
ideal for the imaging of the R Aqr jet. In mid August 2014 the visual
magnitude was about 1 mag brighter. At maximum the system reaches 
a visual brightness of $m_{\rm vis}\approx 6$~mag. 


\begin{figure}
\includegraphics[trim=2cm 13cm 2cm 2cm,clip, width=12cm]{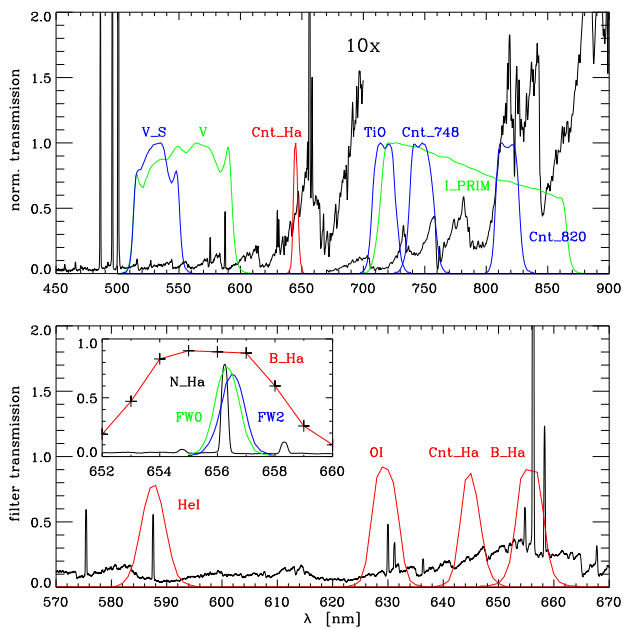}
\caption{R Aqr spectrum from 2010/2011 and transmission curves of the used 
ZIMPOL continuum filters (upper panel) and line filters (lower panel). 
The flux scale for the R Aqr spectrum is arbitrary, but the
short wavelength part $\lambda<700$~nm in the upper panel is multiplied
by a factor of ten for better visibility.}
\label{LineFilters}
\end{figure}

The selected filters cover a wide 
wavelength range using the broad band V-filter where R Aqr is faint, and 
narrow band filters at longer wavelength where the system is bright.
The filter passbands are shown in Fig.~\ref{LineFilters}
together with R Aqr spectra taken by Christian Buil
in August~2010 (spectrum $\lambda= 670 - 900$~nm) and 
September~2011 ($\lambda<700$~nm and emission line spectrum) during 
similar brightness phases of R Aqr as our SPHERE data. These
spectra are available in the database of ARAS (Astronomical Ring for Access
to Spectroscopy)\footnote{Website: www.astrosurf.com/aras.}.
We took also broad-band I\_PRIM data for an investigation of the
dynamic range of the polarimetric mode which will be discussed
in a future paper. These data are either
saturated or taken with a coronagraph and therefore not useful
for photometry. 

The emission nebula of R Aqr was observed with
all ZIMPOL line filters, in particular the different types
and combination possibilities of the H$\alpha$ filters. 
The narrow N\_Ha filters with a width of $\approx 1$~nm
are optimized for continuum rejection with the disadvantage that
the transmission changes rapidly for emission which is
slightly shifted in wavelength. It should be noted that
the N\_Ha filter in the common filter wheel FW0 has its peak
at the rest wavelength 656.3~nm of the H$\alpha$ line, while
the N\_Ha filter in filter wheel FW2 is at 656.5~nm. 

In August 2014 all Cnt\_Ha and
B\_Ha filter frames were saturated in the peak of the
red giant. In October 2014 one set of N\_Ha frames 
is taken with a five-point dithering pattern. Also four frames
covering off-axis fields centered about 2 arcsec to the
NE, SE, SW, and NW of R Aqr were taken. 

Observations in polarimetric mode were taken in October 2014.
We use in this work only the total intensity images of 
the polarimetric data. The throughput is about 18~\% 
lower because of the inserted polarimetric 
components. Also the detector mode is changed for on-chip
demodulation of the data \citep[see][]{Schmid12}. In particular,
the slow modulation modes has a low detector gain of
1.5~e$^-$/ct (ct = ADU, analog to digital count units),
much lower than the 10.5~e$^-$/ct gain for imaging and
fast polarimetry. The read-out noise of the CCD 
(charge-coupled device) is about 
1-2~ct and this translates for the slow polarimetry
into a much lower noise level in terms of photo-electrons e$^-$,
and therefore a better faint source sensitivity,  
when compared to imaging and fast polarimetry.

\subsection{ZIMPOL data reduction}

The data reduction of ZIMPOL imaging data is for most steps 
straight forward and follows standard procedures like bias frame 
subtraction, cosmic ray removal with a median filter, 
and flat fielding. The data reduction was performed with the
(SPHERE-ZIMPOL) SZ-software package, which is written in IDL and was 
developed at the ETH (Eidgen\"{o}ssische Technische Hochschule) 
Zurich. The basic procedures are essentially identical to the 
SPHERE DRH-software provided by ESO.  

A special characteristic of the ZIMPOL detectors are the row masks
which cover every second row of the detector. This feature is
implemented in ZIMPOL for high precision imaging polarimetry using
a modulation-demodulation technique \citep{Schmid12}. 
A raw frame taken in imaging mode has only every 
second row illuminated and the useful science data 
has a format of $512 \times 1024$ pixels consisting of the  
512 illuminated rows of the $1024 \times 1024$ pixel 
detector. One pixel represents $7.2 \times 3.6$ mas on the
sky because of the cylindrical micro lens arrays on the
ZIMPOL detectors, and the total image
covers about $3.6''\times 3.6''$ \citep[see][]{Schmid12}. 

The same image format results from polarimetric imaging. 
In polarimetry, the photoelectric charges 
are shifted up and down by one row during the integration, 
synchronously with the polarimetric modulation. In this way the
whole detector array is filled with photo electrons with
perpendicular and parallel polarization signals stored in 
the ``even'' and ``odd'' rows, respectively. However, the 
image sampling is identical to the imaging mode with photons 
only detected at the position of the 
open detector rows. In this work, only the intensity signal from 
the polarimetric observations is considered and the ``even'' and
``odd'' row counts are added to yield total intensity images with 
$512 \times 1024$ pixels. 

In one of the final steps in the data processing,
the $512 \times 1024$ pixel images are 
expanded with a flux-conserving linear interpolation 
to yield a square $1024 \times 1024$ pixel image where one pixel represents
$3.6 \times 3.6$ mas on sky. The artificial oversampling of the image
rows has no significant apparent effect on the resulting images.  

The detector dark current was found to be variable for our test 
observations. Therefore, it was not possible to subtract the
dark current based on calibration measurements. As an alternative, we 
determined the dark current from the science frames. In imaging mode
one can use, for dark current estimates, the count level in the covered 
row pixels for detector regions with low illumination, and for 
polarimetric observations we used the count level in the detector edges 
(far from the central source) for weakly illuminated frames.  

The estimated dark current levels are given in Table~\ref{RAqrdata}.
Typically the values are small except for longer integrations
with the low gain slow polarimetry mode. Nonetheless, a subtraction
of a low dark current level of $< 1$~ct/pix is still important
for flux measurements in large apertures, for example, $r_{\rm ap}>0.1''$ 
with $\gg 10^3$ pixels, as described in this study for certain cases. 

\begin{table*}
\caption{PSF parameters for the red giant in R Aqr and the standard 
star HD 183143 in different filters.}  
\label{TabPSF}
\begin{tabular}{lclccrccccccc}
\noalign{\hrule\smallskip}
filter & $\lambda_c$ & files / cam  & $\tau_0$ & Strehl $S_0$  
           & \multispan{7}{\hfil${\rm ct}(r)/{\rm ct}_{\rm 3dia}$ [\%]\hfil} 
           & ct$_{\rm 1M}/{\rm ct}_{\rm 3dia}$ \cr
       & [nm]   &         & [ms]  & ratio  &  $r$ [pix] =\hspace{-0.5cm}
       & 0      &    5    & 10   & 30     & 100    & 300   & [\%] \cr  
       &        &         &      & [\%]   & npix =\hspace{-0.5cm}
       &  1        &    81   & 317  & 2821   & 31417  & 282697 \cr
\noalign{\smallskip\hrule\smallskip}
\noalign{\noindent R Aqr, OBS284\_xx \smallskip}
V      & 554   &  0055-58 / 1+2  & 3.3   & 1.9\tablefootmark{a}  &
       & 0.091 &  5.0    & 10.5   & 18.8   & 42.7  & 86.9  & 114  \cr
CntHa  & 645   &  0051-54 / ~~1  & 2.9   & 4.2\tablefootmark{a}     &
       & 0.151 &  8.6   & 18.2   & 28.7  & 45.3   & 86.1   & 118  \cr 
~~{\it max} &  & {\it 0051 / ~~1}      &       & {\it 5.9\tablefootmark{a}}     & 
       & {\it 0.215}     & {\it 10.5} & {\it 19.3} 
             & {\it 28.7}  & {\it 45.8}   & {\it 86.8}     \cr
~~{\it min} &  & {\it 0053 / ~~1}      &       & {\it 3.2\tablefootmark{a}}     & 
       & {\it 0.114}     & {\it 7.0}  & {\it 16.0} & 
            {\it 26.7}  & {\it 43.2}   & {\it 85.2 }     \cr
TiO    & 717   &  0043-46 / ~~1  & 2.8   & 8.3\tablefootmark{a}     &
       & 0.244     &  12.5  & 23.4  & 35.1  & 50.2  & 89.9 & 102 \cr
Cnt748 & 747   &  0043-46 / ~~2  & 2.8   & 10.3\tablefootmark{a}     &
       & 0.280    &  14.1  & 25.7  & 38.3  & 52.6  & 91.0  & 106 \cr
Cnt820 & 817   &  0047-50 / 1+2  & 2.7   & 23.9\tablefootmark{a}     &
       & 0.541     & 25.4  & 41.0  & 54.9  & 65.3  & 93.9  & 103 \cr
\noalign{\smallskip}
\noalign{\noindent HD 183143, STD261\_xx \smallskip}
V      & 554   &  0017-20 / 1+2  & 8.8    & 8.8  &
       & 0.431 &  18.9   & 31.9  & 43.6  & 65.9  & 93.6  & 106  \cr
N\_R   & 646   &  0013-16 / 1+2  & 8.1    & 13.5  &
       & 0.489 &  21.1   & 35.9  & 48.9  & 66.9  & 94.4  & 105  \cr
N\_I   & 790   &  0021-24 / 1+2  & 8.1    & 20.9   &
       & 0.505 &  24.4   & 39.9  & 54.9  & 69.2  & 94.7  & 105  \cr
\noalign{\smallskip\hrule\smallskip}
\end{tabular}
\tablefoot{PSF parameters are coherence time $\tau_0$, 
approximate Strehl ratios $S_0$, and encircles counts 
for different wavelength bands. Encircled counts Ct$(r)$ are given 
for round synthetic apertures with radius $r$ and total number of 
pixels npix and expressed as ratio ${\rm ct}(r)/{\rm ct}_{3dia}$ 
relative to the counts ${\rm ct}_{\rm 3dia}$ in an aperture with a diameter
of $3''$ or $r=416$~pix (npix = 543657). The ratio in the last
column compares the counts on the full 1k$\times$1k area of the
detector with the counts in the round 3$''$-aperture.\\
\tablefoottext{a}{R Aqr is a source with strong (intrinsic circumstellar) 
scattering and therefore this is not a Strehl ratio of a point source.}
}
\end{table*}

\subsection{SPHERE adaptive optics performance 
for the R Aqr observations} 
\label{AOperform}
\begin{figure}
\includegraphics[trim=2cm 12.5cm 3cm 5cm,clip, width=12cm]{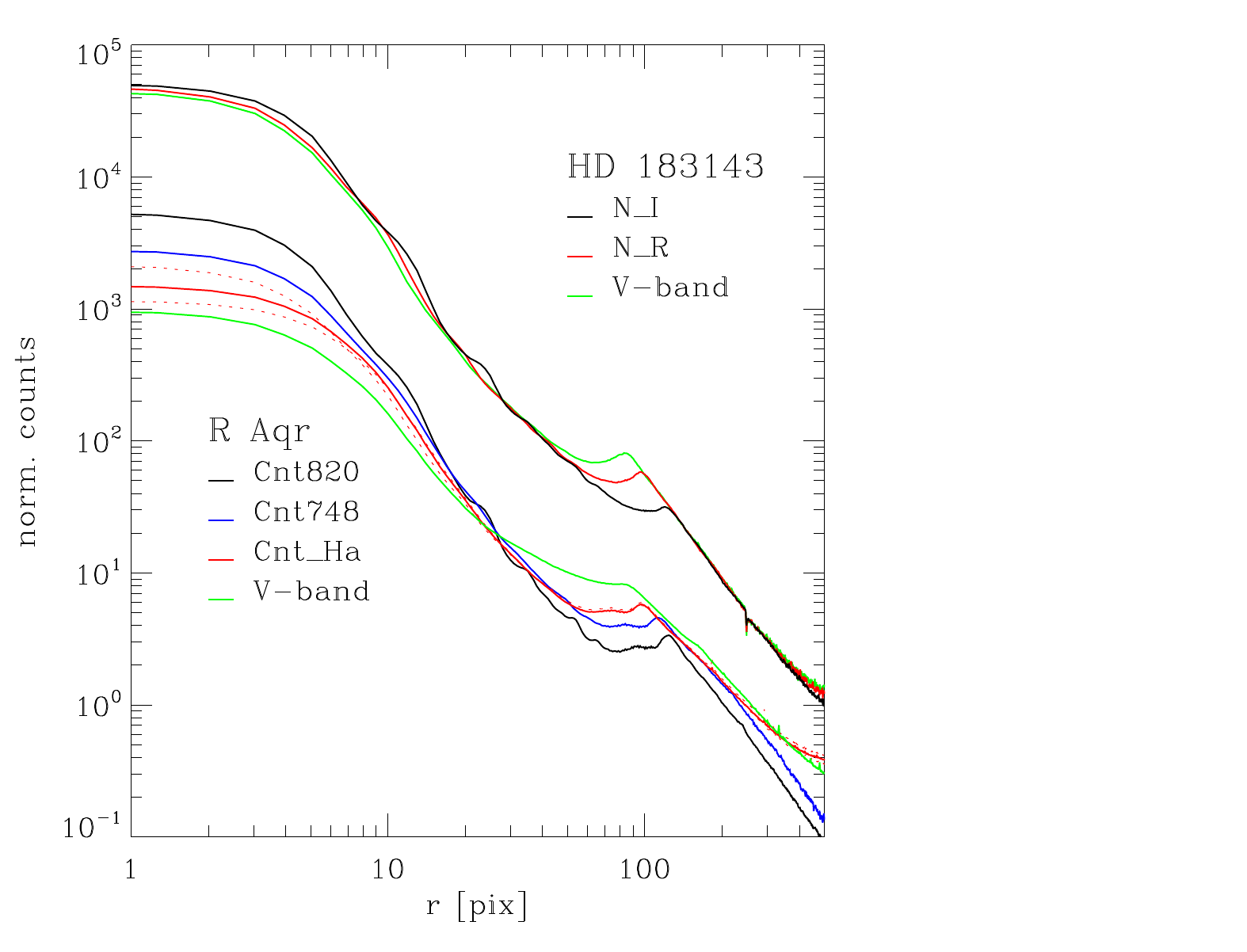}
\caption{PSFs in different filters
for the red giant in R Aqr and the star HD 183143. 
The PSFs are normalized for R Aqr to a total count level of $10^6$ 
within an aperture of 3 arcsec diameter (lower curves) and for HD 183143 
to $10^7$ (upper curves). PSF for different filters are given by
different colors as indicated. The dashed red curves for R Aqr 
show the ``best'' (max) and ``worst'' (min) PSF for the CntHa-filter 
observations.}
\label{RAqrPSF}
\end{figure}

Knowledge of the AO performance and the point spread function (PSF) 
is important for quantitative photometric measurements from 
imaging data. No systematic study on the AO 
performance of SPHERE-ZIMPOL exists up to now. Therefore,
we compare the PSFs for R Aqr observations 
with the PSFs of \object{HD 183143} from
the ESO archive (frame: STD261\_0013-24 from 2015)
which were taken under very good atmospheric conditions.
HD 183143 ($m_V=6.9^m$) is a 
high polarization standard star
which was observed for 
polarimetric calibrations and as a PSF test source. 
The azimuthally averaged and normalized PSFs of R Aqr and 
HD 183143 are plotted in Fig.~\ref{RAqrPSF} and 
approximate Strehl ratios and counts within 
circular apertures are given in Table~\ref{TabPSF}.

For \object{HD 183143} the radial profiles are given for 
the V-, N\_R-, and N\_I-band filters and the total counts
are normalized to $10^7/{\rm ct}_{\rm 3dia}$, where ${\rm ct}_{\rm 3dia}$ are 
the counts within an aperture with a diameter of 3~arcsec. 
The PSFs show clearly the wavelength dependent location of the 
AO control radius at $\lambda/D=20$ 
(at $r\approx 100$~pix in Fig.~\ref{RAqrPSF})
up to which the AO corrects for the wave-front aberrations. 
Beside this feature the PSFs 
for the different filters are quite 
similar with only a very small wavelength dependence in the 
relative peak counts ${\rm ct(0)}/{\rm ct}_{\rm 3dia}$. 

Approximate Strehl ratios $S_0$ are derived from the 
ratio between the measured relative 
peak counts ${\rm ct(0)}/{\rm ct}_{\rm 3dia}$ and the 
expected relative peak flux for diffraction-limited PSFs according to 
\begin{displaymath}
S_0 = {{\rm ct(0)/ct}_{\rm 3dia}\over f(0)/f_{\rm 3dia}}\,,
\end{displaymath} 
where $f(0)/f_{\rm 3dia}$ is calculated
for the VLT (8.0~m primary mirror telescope with a 1.1~m central
obscuration by the secondary mirror). For a pixel size of 
$3.6 \times 3.6$ mas there is $f(0)/f_{\rm 3dia}=4.95~\%$, 3.66~\%, 
and 2.44~\% for the V-, N\_R-, and N\_I-filters, respectively. 
Ratios $S_0$ between about 
9~\% and 21~\% are obtained for the HD 183143 observations 
which were taken under good conditions with a long
atmospheric coherence time of $\tau_0\approx 8-9$~ms. 

This kind of approximate Strehl ratio $S_0$ determination provides
a very useful parameter for a simple comparison of the 
atmospheric conditions and the AO performance 
of different data sets taken with ZIMPOL. However, the $S_0$ 
value does not describe well the SPHERE AO system, 
because of instrumental effects not related to the adaptive optics.
A more sophisticated AO characterization should
be based on the analysis 
of the Fourier transform of the aberrated image as described 
in \citet{Sauvage07}. Such an analysis yields for the N\_I-filter 
PSF of HD 183143 an AO Strehl ratio of 33~\% instead of the 22~\% 
indicated in Table~\ref{TabPSF}. The difference can be explained by a 
residual background at low 
spatial frequencies of undefined nature, perhaps due to 
instrumental stray light, not fully corrected detector noise, 
or other effects. 

For R Aqr the PSFs for the V-band, CntHa, Cnt748, and Cnt820 
filters are plotted in Fig.~\ref{RAqrPSF} and corresponding $S_0$-values
and relative encircled fluxes  
are listed in Table~\ref{TabPSF}. These profiles were normalized to
the total aperture count of ${\rm ct}_{\rm 3dia}=10^6$ to displace them
in Fig.~\ref{RAqrPSF} from the curves of HD 183143. 


The PSFs for R Aqr show a very strong and systematic wavelength 
dependence in the normalized peak counts ${\rm ct(0)/ct}_{\rm 3dia}$
(Table~\ref{TabPSF}, Fig.~\ref{RAqrPSF}). 
For the Cnt820-filter, this value is comparable to the case of 
HD 183143, but for the V-band filter it is about a factor of five lower
than for the Cnt820-filter in R Aqr or the V-band data
of HD 183143. There are several reasons for the lower
PSF peak in R Aqr at shorter wavelengths: (i) the AO performance
was less good than for HD 183143 because the coherence
time was shorter $\tau_0\approx 3$~ms and this affects mainly
the short wavelengths, (ii) the extreme red color of R Aqr could be 
responsible, because the SPHERE wave-front 
sensor ``sees'' essential only I-band light and the AO system 
cannot correct for additional (differential) aberrations in the
V-band, and (iii) the intrinsic PSF of R Aqr is more
extended at short wavelength due to scattering
of the light by circumstellar dust, some light contributions
from the hot binary companion, and/or some extended nebular
emission. 

The PSFs of R Aqr show also temporal variability which are particularly 
strong at short wavelengths. Such variations can certainly be
associated with strong variability in the atmospheric conditions and 
the resulting AO performance. We have picked the case of the four
CntHa-filter exposures from October 11 for which Fig.~\ref{RAqrPSF}
and Table~\ref{TabPSF} give also the parameters for the maximum 
(``best'') and the minimum (``worst'') PSF. For example, the $S_0$ Strehl 
ratio changes within a few minutes by a factor of almost two. 

Because of this strong variability it is difficult to derive
accurate photon fluxes for clouds in dense fields which
need to be measured with small apertures. It is also
difficult to quantify the flux of the extended PSF halo, which should
be disentangled from possible diffuse intrinsic emission in 
the R Aqr system. This problem is taken into account in 
Sect.~\ref{SectHaCloudPhot} for the H$\alpha$ flux measurements 
of the jet clouds.

\begin{figure}
\includegraphics[width=8.8cm]{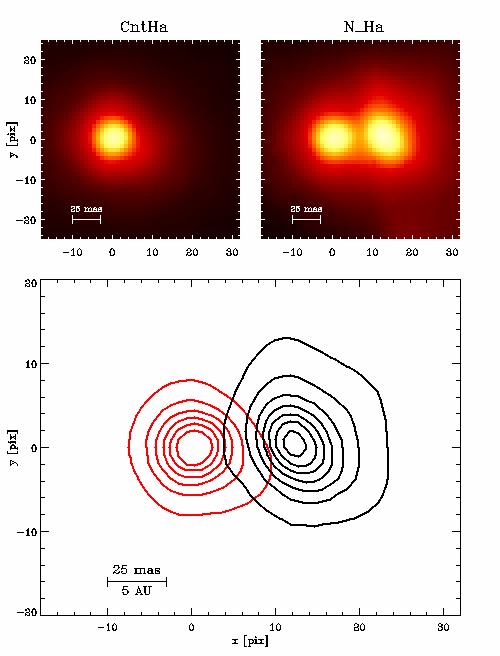}
\caption{R Aqr central binary as seen in the 
CntHa (top left) and the N\_Ha (top right) filters on October 11, 2014
(OBS284\_0051). The contour plot shows the red giant
for the CntHa filter (red) and the "pure" H$\alpha$ emission
in the N\_Ha image after subtraction of the scaled and aligned 
CntHa frame (black). Contour levels are given for seven levels from 
1000 to 7000 ct/pix.}
\label{FigBinary}
\end{figure}

\section{Astrometry of the central binary star system}
\label{SectBinary}

\begin{figure}
\includegraphics[trim=2cm 13cm 2cm 3cm,clip,width=8.5cm]{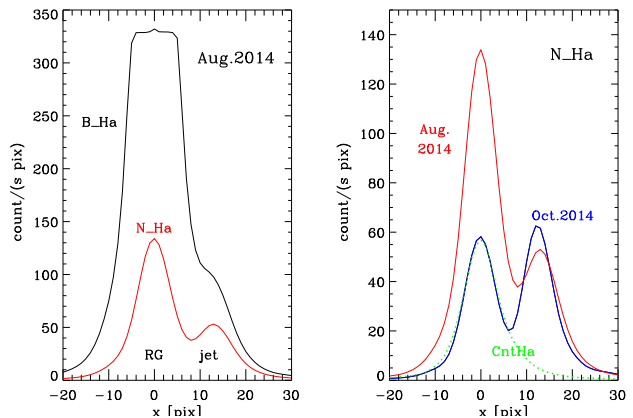}
\caption{East-west cuts through the R Aqr binary system for different 
H$\alpha$ filter observations taken in imaging mode. 
Left: B\_Ha and N\_Ha profiles from
August 12, 2014, showing the much reduced continuum throughput in
N\_Ha (the peak of B\_Ha is strongly saturated); Right: N\_Ha profiles
for August 12, 2014 and October 11, 2014  
illustrating the brightness change of the 
red giant within 60 days. The dotted green curve shows the scaled 
red giant profile as seen in the CntHa filter (slow pol. mode)
from October 11.}
\label{EWcut}
\end{figure}

The R Aqr images taken with SPHERE-ZIMPOL show in all filters, 
except the H$\alpha$ filters, one strongly dominating point source 
from the mira variable. Contrary to this, all H$\alpha$ 
images show in the center two point-like 
sources (Fig.~\ref{FigBinary}) and additional weaker emission features in an
NE and SW direction. The central H$\alpha$ source has a peak flux which
is more than ten times stronger than all other H$\alpha$ features. The central
H$\alpha$ source is located about 12 -- 13 pixel ($\approx 45$~mas) 
to the W of the mira variable which is the expected binary separation for
the R Aqr system \citep[e.g.,][]{Gromadzki09a}. The central 
H$\alpha$ is slightly extended in the NE--SW jet direction
(Fig.~\ref{FigBinary}b and c). Figure~\ref{EWcut} shows
E-W profiles through the two source peaks for the different
H$\alpha$ filter observations.  

The bright point-like H$\alpha$ source, just besides the mira 
in R Aqr, is very likely the compact emission region around
the active companion, which is at the same time the jet source of
the system. Thus, our high resolution H$\alpha$ images provide the
unique opportunity for an accurate measurement of the binary separation 
and orientation for the R Aqr system, and with future observations
it should be possible to determine 
accurate orbital parameters and stellar masses. 

Ideal for accurate astrometric measurements are the
N\_Ha data from October 11, 2014 in which the two stellar sources
have about the same brightness. For this date the mira was at its
minimum phase, and about a factor 2.4 fainter 
(peak intensity) than in the N\_Ha data of
August 12. In the B\_Ha images of the same date, the mira variable 
is much brighter and saturated because of the wider band pass
(see Fig.~\ref{EWcut}).

The simultaneous observations in the Cnt\_Ha in camera 1 
and the N\_Ha in camera 2 from October 11 (OBS28\_0051-54) provide
as an additional advantage a binary image and a ``reference PSF''
for the mira with the same PSF distortions.  
From this set we selected the ``best'' or maximum PSF
exposure (OBS284\_0051, Table~\ref{TabPSF}). We centered the CntHa PSF 
of the red giant and used it as astrometric zero point 
(Fig.~\ref{FigBinary}a). Then we subtracted a scaled version of this frame 
from the N\_Ha double star image (Fig.~\ref{FigBinary}b), which was 
shifted around in steps of 0.1 pixels in $\Delta x$ and $\Delta y$ 
until the subtraction yields ``a clean H$\alpha$'' 
source image (Fig.~\ref{FigBinary}c)
with a minimal residual pattern at the zero point, that is, at the 
location of the subtracted red giant PSF. 
A centroid fit to the H$\alpha$ source center yields then
the following relative position of the H$\alpha$
source with respect to the red giant:
\begin{eqnarray}
\Delta x = 12.54\pm 0.10~{\rm pix}\,,
        &\quad &  \Delta y=0.54\pm0.10~{\rm pix}\,,\\
\Delta\alpha = -45.1\pm 0.6~{\rm mas} \,,&\quad & 
        \Delta\delta = 0.4\pm 0.6~{\rm mas} \,.  
\end{eqnarray} 
This corresponds with the astrometric calibration given in
Sect.~\ref{SectZimpol} to a position angle of $270.5^\circ\pm 0.8$ 
(measured N over E) and a separation of $45.1\pm 0.6$ mas. This
translates for an R Aqr distance 
of 218 pc \citep{Min14} into an apparent separation of 9.8~AU. 

It is unclear how well this astrometric result for
the photo-centers represents the positions of the mass 
centers of the two stellar components. 
The mira variable shows a photosphere 
with an asymmetric light distribution \citep{Ragland08} and for 
the jet source it seems likely that the measured H$\alpha$ emission 
peak is not exactly at the position of the invisible
stellar source probably located in an accretion disk. 

Therefore, the position of the photo-center could 
deviate from the mass center of the stellar
components by more than the indicated photo-center 
measurement uncertainties of $\approx 0.5$~mas. Future 
observations will show how well one can 
determine the orbit from the photo-center measurements. In any case
we can expect a significant reduction of uncertainties
for the orbital parameters of R Aqr.

\begin{table*}
\caption{R Aqr red giant continuum photometry for an aperture of 
$3.6''\times 3.6''$.}  
\label{PhotRAqr}
\begin{tabular}{lllcccccc}
\noalign{\smallskip\hrule\smallskip}
filter  & files / cam & mode & ct$_{\rm 1M}$  & ct/s  
        & ${\rm am}\cdot k_1$ & $m_{\rm mode}$ & $zp_{\rm ima}$ 
        & $m(F)$ \\ 
        &         &      & [$10^6$]           
        & [$10^6$/s] & [mag]  & [mag] & [mag] & [mag] \\
\noalign{\smallskip\hrule\smallskip}
\noalign{\smallskip date: 2014-08-12, OBS224\_xx \smallskip} 
V\_S    & 0095 / 1+2 & imaging & $1.38\pm 0.06$ & 0.0688     
        &  0.19 & 0.0     &  22.72  & $10.4\pm 0.1$ \\ 
HeI     & 0097 / 1+2 & imaging & $5.27\pm 0.15$ & 0.0527     
        &  0.18 & 0.0     &  20.77  & $8.8\pm 0.1$ \\     
OI      & 0096 / 1+2 & imaging & $5.32\pm 0.15$ & 0.0531     
        &  0.15  & 0.0      &  20.67  & $8.7\pm 0.1$ \\     

\noalign{\smallskip date: 2014-10-11, OBS284\_xx \smallskip} 
V       & 0055-58 / 1+2 & slow pol. & $5.55\pm 0.24$  & 0.551      
        & 0.15  & -1.93  &  23.70  & $11.3\pm 0.1$ \\ 
V       & 0039-42 / 1+2 & fast pol. & $0.073\pm 0.034$ & 0.058      
        & 0.15  & 0.18  &  23.70    & $11.5\pm 0.3$ \\ 
Cnt\_Ha & 0051-54 / ~~1  & slow pol. & $13.1\pm 0.6$  & 0.262      
        & 0.10  & -1.93  & 20.35    & $8.8\pm 0.1$ \\ 
TiO\_717& 0043-46 / ~~1 & fast pol. & $0.307\pm 0.021$ & 0.244      
        & 0.09  & 0.18   & 21.78    & $8.0\pm 0.1$ \\ 
Cnt748  & 0043-46 / ~~2 & fast pol. & $1.05\pm 0.03$   & 0.836      
        & 0.08  & 0.18   & 21.68    & $6.6\pm 0.1$ \\ 
Cnt820  & 0047-50 / 1+2 & fast pol. & $4.16\pm 0.09$    & 3.31       
        & 0.08  & 0.18  & 20.97     & $4.4\pm 0.1$ \\ 
\noalign{\smallskip\hrule\smallskip}
\end{tabular}
\tablefoot{Column~4 lists the dark corrected 
counts per frame $ct_{\rm 1M}$ with measuring errors, ct/s are the
count rates, ${\rm am}\cdot k_1$ the atmospheric extinction correction,
$m_{\rm mode}$ the mode dependent transmission offset, $zp_{\rm ima}$ the
photometric zero point for the imaging mode, and $m(F)$ the resulting filter
magnitudes.}
\end{table*}

\section{ZIMPOL aperture photometry}

\subsection{Aperture photometry for the red giant}
\label{APhot}
\object{R Aqr} shows strong, periodic brightness variations between 
$m_{\rm vis}\approx 6.5^m$ and 11.0$^m$. 
We derive photometric magnitudes of the mira variable for our ZIMPOL 
filter observation which are useful for the absolute H$\alpha$ 
line fluxes of the jet clouds, the determination of upper flux limits
for the hot companion, and for estimates of the flux contribution of 
the mira star to the HST line filter images. Photometric magnitudes are obtained by summing up all counts
``ct$_{\rm 1M}$'' registered in the $10^6$ pixels area 
[$x_1$:$x_2$,$y_1$:$y_2$] = [13:1012,13:1012] of the
$1024\times 1024$ pixel detector. This is equivalent to 
photometry with an aperture of $3.6'' \times 3.6''$ using
essentially the whole detector except for the outermost rows and columns 
of the CCD, which are partly hidden by the frame 
holding the microlens array of the detector \citep[see][]{Schmid12}.  

The obtained counts ${\rm ct}_{\rm 1M}$ per frame
and detector arm are listed in column 4 of Table~\ref{PhotRAqr}. 
The indicated measuring uncertainty is composed of three
error sources; a relative factor of 
$\pm 0.02\cdot {\rm ct}_{\rm 1M}$ which accounts 
for sky transparency and instrument throughput variations,
a bias subtraction uncertainty of $\pm 20000$~ct equivalent 
of 0.02 counts/pixel, and a relative uncertainty in the dark current 
subtraction of $\pm 20~$\% which becomes more important
than the bias subtraction uncertainty for a dark current $>0.1$ ct/pix.  
These contributions are treated like independent errors and are
combined by the square-root of the sum of the squares. 

Count rates ct/s (Table~\ref{PhotRAqr}) are obtained by dividing 
${\rm ct}_{\rm 1M}$ with the
detector integration time (DIT). A small correction for the frame
transfer time (ftt) is required
${\rm ct/s} = {\rm ct}_{\rm 1M}/({\rm DIT}+{\rm ftt})$
because the detector is also illuminated during the short
frame transfer, which is ${\rm fft}= 56$~ms for imaging and fast polarimetry 
and ${\rm fft}=74$~ms for slow polarimetry \citep{Schmid12}.
The frame transfer time correction is only relevant ($>1~\%$) 
for short DIT $<7$~s. 
 
In the next step the count rates in a given filter $F$
are converted to photometric magnitudes $m(F)$ using the formula
\begin{equation}
m(F) = -2.5\,{\rm log}\,({\rm cts/s}) - {\rm am}\cdot k_1(F)
        -m_{\rm mode} + zp_{\rm ima}(F)\,.
\label{Eqzeropoint}
\end{equation}
This accounts for the atmospheric extinction with
a filter and airmass dependent 
correction ${\rm am}\cdot k_1(F)$ (Table~\ref{PhotRAqr})
using the Paranal
extinction curve from \citet{Patat11}. The airmass for
our R Aqr observations was in the range ${\rm am}=1.31-1.46$ for Aug.~12 and ${\rm am}=1.11-1.15$ for Oct.~11. 

The photometric zero points $zp_{\rm ima}(F)$ for the
individual filters were determined with calibration measurements 
of the spectrophotometric standard star HR 9087 \citep{Hamuy92}, 
which will be described in Schmid et al.~(2016, in prep.). The 
$zp_{\rm ima}(F)$-values apply for the ZIMPOL imaging 
mode with the gray beam-splitter between wave front sensor and ZIMPOL.
For other instrument modes one needs in addition
a throughput offset parameter $m_{\rm mode}$. 
For example, for polarimetry, $m_{\rm mode}=0.18$ accounts for
the reduced transmission because of the inserted polarimetric components. 
The difference of $\Delta m_{\rm mode} = -2.11$ 
between fast and slow polarimetry is because of the changed detector gain 
factor from 10.5 e$^-$/ct in fast polarimetry to 1.5 e$^-$/ct in 
slow polarimetry.   

The zero point values $zp_{\rm ima}(F)$ given in Table~\ref{PhotRAqr}
are preliminary and the estimated uncertainties are about
$\pm 0.10$~mag and this dominates the error in the 
final magnitudes except for the
underexposed V-band (fast polarimetry) observations. 
For such low illumination, the uncertainty of ($\pm 0.02$ ct/pix)
in the bias level is an issue in the ``full detector'' aperture photometry. 

R Aqr was observed with different filters in August and October 2014. 
We may compare the V\_S ($\lambda_c=532$~nm) magnitude $m_{\rm V\_S}=10.4^m$
from August 11 with $m_V=11.4^m$ ($\lambda_c=554$~nm). This gives
a decrease in brightness of about 1~mag within 60 days in good agreement 
with the AAVSO light curve (see Sect.~\ref{SectObs}). 

The obtained magnitudes (Table~\ref{PhotRAqr}) indicate 
for October 11, 2014 very red colors 
of V -- CntHa = 2.5$^m$ and  V -- Cnt820 = 6.9$^m$ for R Aqr.
Because R Aqr is strongly variable, these colors cannot be
compared readily with literature values. \citet{Celis82} 
measures also a very red color of V -- I = 7.8$^m$ 
in Johnson filters for the R Aqr minimum epoch from Nov. 1981.

\begin{table}
\caption{H$\alpha$ absolute ``aperture'' photometry for R Aqr. } 
\label{HalphaAperture}
\begin{tabular}{llcc}
\noalign{\smallskip\hrule\smallskip}
flux   & ct$_{\rm 1M}$/s            & flux \\
  component            & [kct/s]  &      \\

\noalign{\smallskip\hrule\smallskip}
\noalign{\smallskip date: 2014-08-12, OBS224\_0092+93 / 1+2 \smallskip} 
B\_Ha total        & $>228$         \\
~~~~H$\alpha$ only\tablefootmark{a} 
                   & $44\pm 20$  & $35 (\pm 16) \cdot 10^{-12}$ \\
~~~~RG only\tablefootmark{b}         
                   & $>184$      & $>2.4 \cdot 10^{-12}$ \\
\noalign{\smallskip date: 2014-08-12, OBS224\_0094 / 2 \smallskip} 
N\_Ha (FW2)        & $81\pm 5 $    \\
~~~~H$\alpha$ only\tablefootmark{a}  
                   & $48\pm 17 $ & $64 (\pm  23) \cdot 10^{-12}$ \\
~~~~RG only\tablefootmark{b}         
                   & $33\pm 17 $ & $3.6 (\pm 1.8) \cdot 10^{-12}$ \\
\noalign{\smallskip}
\noalign{\smallskip date: 2014-10-11, OBS284\_0030-34 / 1+2  \smallskip} 
N\_Ha (FW0)        & $63\pm 7$ \\
~~~~H$\alpha$ only\tablefootmark{a}  
                   & $42\pm 12$ & $38 (\pm 10) \cdot 10^{-12}$ \\
~~~~RG only\tablefootmark{b}         
                   & $21\pm 12$ & $2.2 (\pm 1.1) \cdot 10^{-12}$  \\
\noalign{\smallskip date: 2014-10-11, OBS284\_0051 / 2 \smallskip} 
N\_Ha (FW2)        & $330\pm 16$ \\
~~~~H$\alpha$ only\tablefootmark{a}  
                   & $256\pm 25$ & $43 (\pm 10)\cdot 10^{-12}$ \\
~~~~RG only\tablefootmark{b}         
                   & $74\pm 25$  & $1.4 (\pm 0.4)\cdot 10^{-12}$ \\
\noalign{\smallskip weighted mean \smallskip}
~~~~H$\alpha$ only\tablefootmark{b}  
                   & $F_{\rm 1M}({\rm H}\alpha)$ & $42 (\pm 7)\cdot 10^{-12}$   \\
\noalign{\smallskip\hrule\smallskip}
\end{tabular}
\tablefoot{The second column gives the total count rates 
in H$\alpha$ filters, and the count splitting between H$\alpha$ line counts
and red giant continuum counts. The third column lists 
the corresponding H$\alpha$ line flux and red giant continuum flux and,
at the bottom, the weighted mean for the H$\alpha$ flux. \\
\tablefoottext{a}{line flux in units of ${\rm erg}/({\rm cm}^{2}{\rm s})$,}
\tablefoottext{b}{continuum flux in unit of 
${\rm erg}/({\rm cm}^{2}{\rm \AA \,s})$}.
}
\end{table}

\subsection{Total H$\alpha$ flux within $3.6''\times 3.6''$}
\label{Haaperture}

Absolute photometry is required for the determination of
the H$\alpha$ flux of jet clouds for the determination of intrinsic
line emissivities $\epsilon({\rm H}\alpha)$. A first complicating
factor is that the bright red giant is a strongly variable source and
therefore a bad flux reference source. A second complicating
factor are the complex structures of the H$\alpha$
emission features, composed of small and large, bright and faint 
clouds, and perhaps even a diffuse emission component. The flux
measurements require therefore the definition of flux apertures
and appropriate aperture correction factors, and this introduces
additional measuring uncertainties. The quasi simultaneous HST
data are very helpful to improve and check the ZIMPOL flux measurements 
but in the HST data some clouds are blended and 
aperture correction factors differ because of the lower spatial 
resolution.   

For this reason we determine here also the total H$\alpha$ flux for 
the central $3.6''\times 3.6''$ region of R Aqr. This value is
independent of the flux aperture definition and correction factors, 
but needs to account properly for the contribution of the red
giant. Therefore, this provides an alternative H$\alpha$ 
flux comparison with the HST data. The total H$\alpha$ flux in
a large aperture allows also a flux comparison with seeing-limited
spectrophotometric measurements. 

We have taken different H$\alpha$ filter observations 
of R Aqr, simultaneous CntHa / B\_Ha filter data on August~12, 
simultaneous CntHa / N\_Ha(FW2) data on August~12 and October~11 and
N\_Ha(FW0) data in both channels on August~12. The total frame count 
rates ct$_{\rm 1M}$/s are given in Table~\ref{HalphaAperture}.
Two steps are required for the conversion of these measurements into 
a total H$\alpha$ flux for the innermost $3.6''\times 3.6''$ R Aqr 
nebulosity: (i) the emission of the
red giant must be subtracted from the extended 
H$\alpha$ emission, and (ii) the conversion of the count rates 
into a continuum flux for the red giant and a line flux for H$\alpha$
using photometric zero points. 

Table~\ref{HalphaAperture} splits the total count rates into count rates
for the H$\alpha$ line and red giant continuum. 
This is achieved with 
a subtraction of the scaled CntHa image from OBS284\_0051 as 
demonstrated for our astrometry of the 
central binary in Sect.~\ref{SectBinary}. This H$\alpha$
measuring procedure is accurate ($\approx \pm 10~\%$) for the 
simultaneous N\_Ha frame (OBS284\_0051). 

For all other H$\alpha$ observations, there exist no simultaneous,
unsaturated CntHa frames for the flux splitting. Subtracting non-simultaneous
CntHa frames from H$\alpha$ data is less accurate because of the 
AO performance variations described in Sect.~\ref{AOperform}. 
Matching the PSF peak of the red giant might not account well for the 
total red giant flux in the H$\alpha$ image, which is mostly ($> 80~\%$) 
contained in the extended halos at $r>10$~pix.
Therefore the flux splitting is less accurate ($\approx \pm 25~\%$)
for the OBS284\_0030-34 data taken 15 minutes before
the red giant PSF, and the uncertainty is substantial 
($\approx \pm 35-50~\%$) for the H$\alpha$ flux 
from August~12 because then the red giant was brighter and even 
saturated in the B\_Ha data.     

The conversion of the count rates   
${\rm ct}_{\rm 1M}/{\rm s}$ measured in a given H$\alpha$-filter ``$F$'' 
into an emission line flux 
$f({\rm H}\alpha)$ [erg cm$^{-1}$s$^{-1}$]
or into a red giant continuum flux
$f(F)$ [erg cm$^{-1}$s$^{-1}{\rm \AA}^{-1}$] follows from the following 
formula
\begin{equation}
f = {\rm ct_{\rm 1M}/s}\cdot 10^{\,0.4\,(am\cdot k_1+m_{\rm mode})} 
\cdot c_{\rm zp}(F)\,,
\label{Eqctflux}
\end{equation} 
where $c_{\rm zp}(F)$ is either the photometric zero point for the
H$\alpha$ line emission $c_{\rm zp}^\ell(F)$  
or for the continuum emission $c_{\rm zp}^{\rm cont}(F)$. 
The zero point values $c_{\rm zp}$ for the different H$\alpha$ 
filters are given in Table~\ref{HalphaFilters}. They are 
based on calibration measurements
of the Vega-like spectrophotometric standard star HR 9087 for which the 
stellar H$\alpha$ absorption has been taken into account. A detailed 
description of the ZIMPOL filter zero point determination is planned
for a future paper. The atmospheric extinction correction is 
$0.10^m\pm 0.01^m$ and $m_{\rm mode}$-values are $0^m$ for imaging, 
$+0.18^m$ for fast polarimetry, and $-1.93^m$ for slow polarimetry 
as described in Sect.~\ref{APhot}.  

The line flux conversion $c_{\rm zp}^\ell$ depends on the
wavelength of the line emission within the 
filter transmission curve. This is a particularly important issue
for the very narrow transmission profiles of the N\_Ha filters
(see Fig.~\ref{LineFilters}).
For example, the transmission in the N\_Ha filters is reduced 
by 25~\% for an offset of $\Delta\lambda=\pm 0.3~{\rm nm}$ 
($\Delta {\rm RV}=\pm 137$ km/s) from $\lambda_{\rm c}$.
The enhanced uncertainties for $c_{\rm zp}^\ell$ for the 
N\_Ha filters in Table~\ref{HalphaFilters} take this problem into 
account. 
Of course, high velocity H$\alpha$ gas with $|RV|\gg 100$~km/s is not
detected with the N\_Ha filters.  

We adopt a heliocentric radial velocity (RV) of $-25$~km/s 
for the H$\alpha$ emission peak of R Aqr, because the measured values lie 
between $0$ and $-50$~km/s \citep[e.g.,][]{vanWinkel93,Solf85} and 
they agree well also with the systemic radial velocity of $-24.9$~km/s 
derived from the radial velocity curve of the mira by  
\citet{Gromadzki09a}.
This yields for our observing dates a geocentric RV of about 
$-30$~km/s and $-5$~km/s for the H$\alpha$ line, or shifts of
less than 0.1~nm with respect to the H$\alpha$ rest wavelength 
in air of 656.28~nm. 
This matches well with the peak transmission wavelengths
$\lambda_c=656.34$ for the N\_Ha filter located in FW0
and therefore the expected transmission is about $T = 0.7$. The match
with the peak wavelength $\lambda_c=656.53$ for the N\_Ha located in FW2
is less good and the expected transmission is only about $T=0.5$.
These values are used for the H$\alpha$ line flux
determinations for the N\_Ha filters given in Table~\ref{HalphaAperture}. 

\begin{table}
\caption{ZIMPOL H$\alpha$ filter parameters for the filter B\_Ha, N\_Ha
(FW0 and FW2) and CntHa.} 
\label{HalphaFilters}
\begin{tabular}{lcccc}
\noalign{\smallskip\hrule\smallskip}
 value [unit]      &  B\_Ha    & N\_Ha    & N\_Ha  & CntHa    \\
                   &           & (FW0)    & (FW2)              \\
\noalign{\smallskip\hrule\smallskip}
$\lambda_{\rm c}$ [nm]
              & 655.6 & 656.34 & 656.53 & 644.9 \\
FWHM [nm]     & 5.5   &  1.15   & 0.97   & 4.1   \\ 
$T_{\rm peak}$  &  0.89  &  0.76  & 0.70  &  0.87     \\
$\Delta\lambda$ [nm] 
              &  5.35  &  0.81   & 0.75   &  3.83   \\
\noalign{\medskip}
$c^\ell_{\rm zp}$ &  7.2 &  8.4 & 9.2 & ...  \\ 
               & $(\pm 0.4)$ & $(^{+2.0}_{-0.5})$ & $(^{+4.0}_{-0.5})$ &     \\
\noalign{\medskip}
$c^{\rm cont}_{\rm zp}$ & 1.20 & 9.3 & 10.0 & 1.59 \\
                 & $(\pm 0.05)$ & $(\pm 0.5)$ 
                             & $(\pm 0.5)$ & $(\pm 0.05)$ \\
\noalign{\smallskip\hrule\smallskip}
\end{tabular}
\tablefoot{Filter parameters are central wavelength 
$\lambda_{\rm c}$, full width at half maximum (FWHM), 
peak filter transmission $T_{\rm peak}$, filter equivalent width 
$\Delta \lambda = \int T(\lambda)\,{\rm d}\lambda$,
photometric zero points for the H$\alpha$ line emission 
$c_{\rm zp}^\ell$ at $\lambda_{\rm c}$ in units of 
$10^{-16}{\rm erg/(cm}^{2}{\rm ct})$ and for 
the continuum emission $c_{\rm zp}^{\rm cont}$ in units of 
$10^{-17}{\rm erg}/({\rm cm}^{2}{\rm \AA\, ct})$. 
The uncertainties for $c^\ell_{\rm zp}$ for N\_Ha filters
consider also H$\alpha$ wavelength shifts of $\pm 0.3$~nm from
$\lambda_{\rm air}({\rm H}\alpha)$.} 
\end{table}

H$\alpha$ wavelength shifts are much less critical for the 
broader, 5~nm wide B\_Ha filters. But the B\_Ha photometry 
could be contaminated by the nebular [N\,II] emission located at
654.8 and 658.3 nm (see Fig.~\ref{LineFilters}). 
For the central region of R Aqr, the H$\alpha$ emission
is about ten times stronger than [N\,II] 
as follows from the spectrum shown in Fig.~\ref{LineFilters} or 
the HST fluxes given in Table~\ref{HSTLinetable}. Table~\ref{HalphaAperture} lists the resulting flux values and we
use in the following the weighted mean value 
$F_{\rm 1M}({\rm H}\alpha)$ for the H$\alpha$ line flux
within the $3.6''\times 3.6''$ central area. Better results could be obtained with simultaneous, non-saturated 
B\_Ha / CntHa measurements, because the broader line filter is less
affected by H$\alpha$ radial velocity shifts. An additional measurement
in the N\_Ha / CntHa filter could be useful, if the disturbing 
continuum emission is strong, or if contamination by [N~II] emission
is an issue. Previous spectroscopic observations by \citet{vanWinkel93} give 
26 and $40\cdot 10^{-12} {\rm erg\,s}^{-1}{\rm cm}^{-2}$ for the
H$\alpha$ flux of R Aqr for two epochs in 1988. Thus the total H$\alpha$ 
flux is on a similar level as 25 years ago.

\section{ZIMPOL photometry for the jet clouds}

\subsection{Overview on the H$\alpha$ jet cloud structure}

\begin{figure*}
\includegraphics[width=19cm]{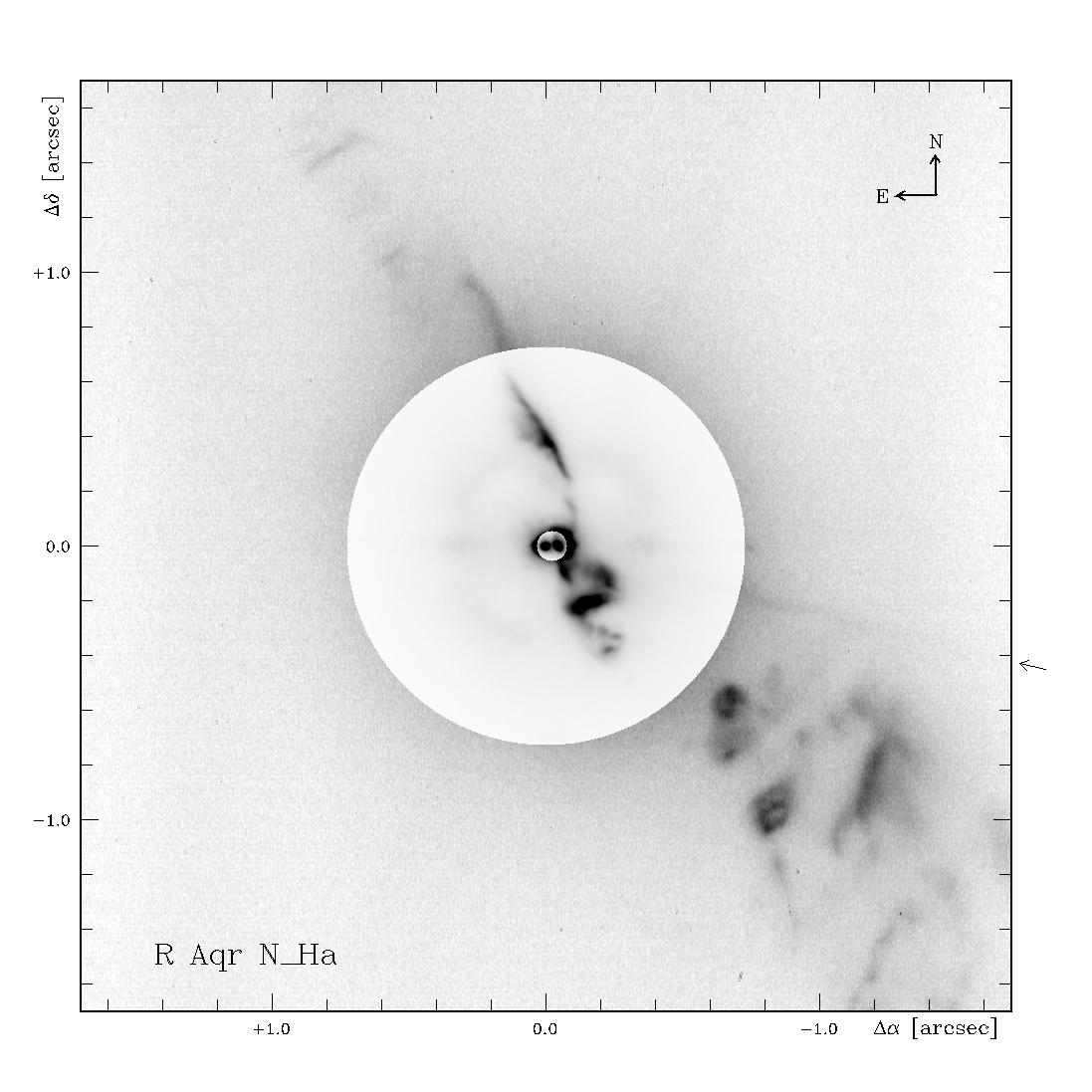}
\caption{Map of the SPHERE-ZIMPOL H$\alpha$ line observation of R Aqr 
taken in the N\_Ha (FW0) filter on Oct. 11, 2014. The field is
divided into three different gray scale regions ranging from 0 to 1000
ct~pix$^{-1}$frame$^{-1}$ for the central binary, 0 to 100 for the inner jet 
region $r_{\rm RG}<0.72''$ , and 0 to 10 for the outer jet region 
$r_{\rm RG}>0.72''$. The red giant is at the zero point of the 
coordinate system. There is a small angle offset of 2 degr of the 
sky N orientation in a counter-clockwise direction with respect the 
(vertical) $y$-direction. A weak instrumental spike from a
telescope spider is present at a position angle of about $-105^\circ$
indicated by a small arrow outside the frame.}
\label{Haoverview}
\end{figure*}
 
\begin{figure}
\includegraphics[width=8.8cm]{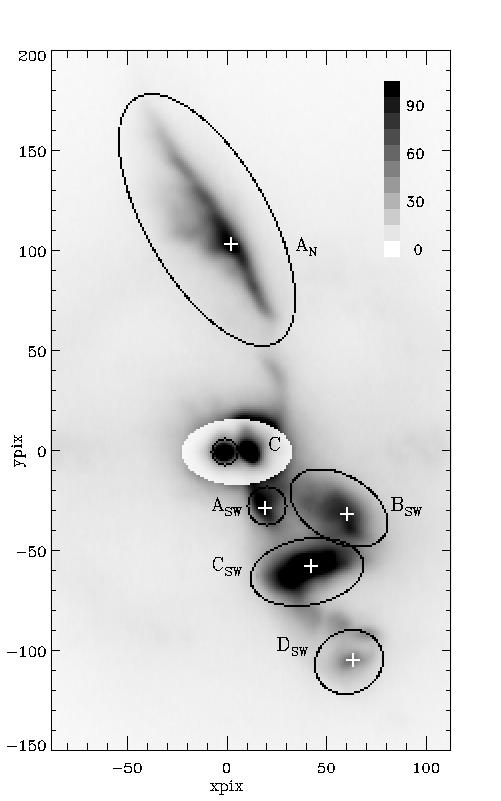}
\caption{R Aqr observation from Oct. 11, 2014 in the N\_Ha filter 
of the inner jet of R Aqr with cloud identification, 
astrometric points (defined for
observations taken on Oct. 11, 2014), and used flux apertures. The red giant
is located at $(x_{\rm pix},y_{\rm pix}) = (0,0)$ and the pixel scale
is 3.60 mas/pix.}
\label{innerjet}
\end{figure}

\begin{figure}
\includegraphics[width=9.5cm]{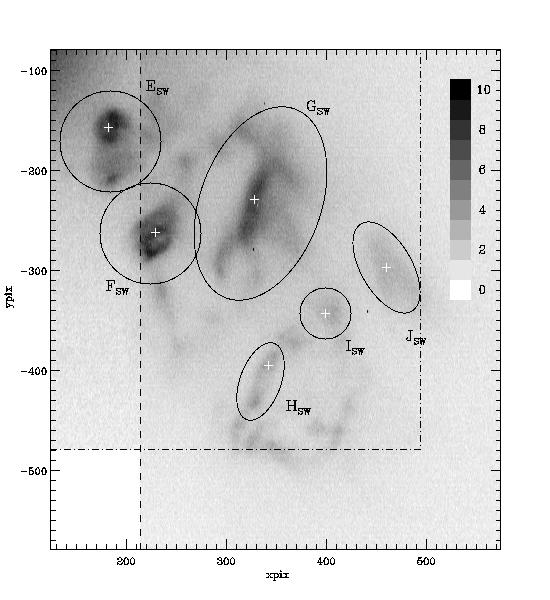}
\caption{Images of the outer bubbles in the SW from Oct. 11, 2014
taken with the N\_Ha-filter. 
The region on the right of the dashed line is from the
off-axis field observation OBS284\_0038 while the narrow section
on the left is from the image shown in Fig.~\ref{Haoverview}, which
covers the region marked with the dashed dotted line. 
No data were taken for the lower left corner region. The red giant
is located at $(x_{\rm pix},y_{\rm pix}) = (0,0)$ and the pixel scale
is 3.60 mas/pix. } 
\label{swbubbles}
\end{figure}

\begin{figure}
\includegraphics[width=8.5cm]{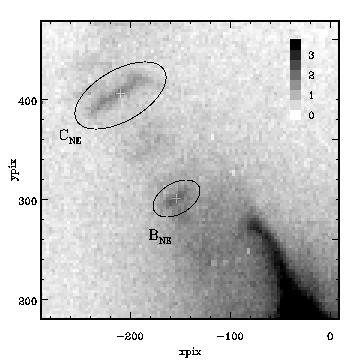}
\caption{N\_Ha-filter image of the NE-jet wisps. The
image shows mean values for $3\times 3$ binned pixels in
order to reduce the noise. The axis indicates the
pixel coordinates for the unbinned frame where (0,0) is
the coordinate of the red giant.}  
\label{NEwisp}
\end{figure}

The H$\alpha$ map plotted in Figure~\ref{Haoverview} shows the 
very rich nebular emission within the central 
$3.4'' \times 3.4''$ region of R Aqr based on the five dithered N\_Ha 
images from camera 1 and camera 2 with a total of 
200 s integration time (OBS284\_0030-34). Individual jet features 
are identified in Figs.~\ref{innerjet}, \ref{swbubbles} ,
and \ref{NEwisp}, where crosses localize flux peaks or the 
apparent ''centers'' of bright extended structures. We use circles or ellipses 
to define 
the apertures for line flux measurements. Table~\ref{JetTable}
gives the position of the crosses and full lengths and widths
for the clouds $\ell_{\rm cl}$, $w_{\rm cl}$ and apertures $\ell_{\rm ap}$, 
$w_{\rm ap}$. Cloud diameters are FWHM values which are measured 
with a precision of roughly $\pm 10~\%$.

\paragraph{The inner jet.}
We call the intermediate brightness H$\alpha$ emissions located
at distances $0.05''<r_{\rm jet}\lapprox 0.70''$ the inner jet. North of the central binary is one almost straight narrow cloud 
($A_{\rm N}$) extending from 
about $r_{\rm jet}=0.2''$ to $0.6''$. A faint arc seems to connect this
filament with the central source, while on the outside there
is a weak extension out to about $1''$ (218 AU) from the source. 
The whole structure looks like a narrow, slightly undulating 
gas filament with an orientation of 25$^\circ$, which is displaced 
towards the west by about $0.1''$ with respect to a strictly radial 
outflow from the jet source. The inner jet towards the SW has a different morphology with
a string of clouds arranged in a double zig-zag pattern
and a location between about 190$^\circ$ and 240$^\circ$ with respect
to the jet source (see also Fig.~\ref{innerjet}).

\paragraph{The SW outer bubbles.}
There is another group of lower surface brightness clouds in the SW 
at separations between $r_{\rm jet}\approx 0.8''-2.0''$. A variety of 
structures can be recognized; bubble-like structures 
E$_{\rm SW}$, F$_{\rm SW}$, elongated clouds G$_{\rm SW}$, J$_{\rm SW}$,
and a shell-like structure including clouds H$_{\rm SW}$ and
I$_{\rm SW}$. The position of these clouds
is confined to a wedge with an opening angle of about $\pm 15^\circ$
centered along a line with an orientation of about 230$^\circ$ degrees. 

The outer bubbles extend out to the extreme SW-corner of
Fig.~\ref{Haoverview}, which corresponds to a separation of $2.3''$
from the source. The off-axis field image
OBS284\_0038 extends the field of view out to $4''$ in a
SW direction. Part of this outer field is shown in Fig.~\ref{swbubbles}
demonstrating that there is no bright cloud outside 
Fig.~\ref{Haoverview}. There is a very weak trace of
a possible extended cloud at $r_{\rm jet}\approx 3.9''$, 
$\theta_{\rm jet}=230^\circ$, outside the region 
in Fig.~\ref{swbubbles}, with a surface 
brightness $<0.003~{\rm ct/(s\,pix})$ significantly fainter than
for H$_{\rm SW}$, I$_{\rm SW}$, or J$_{\rm SW}$. 

\paragraph{The NE wisps.} Feature C$_{\rm NE}$ at 
$r_{\rm jet}=1.65''$ at a position angle of roughly 
$30^\circ$ is a $0.2''$ long, narrow,
straight H$\alpha$ emission with an orientation perpendicular
to the radial jet direction (Fig.~\ref{NEwisp}).
A second, weaker and shorter such wisp (B$_{\rm NE}$) is seen 
at $r_{\rm jet}=1.2''$ with the same orientation indicating 
the possibility of a close relationship between these two
features.

\subsection{Positions for the H$\alpha$ clouds}

The astrometric positions for the clouds for the points marked
with a cross are given in Table~\ref{JetTable} in
polar coordinates $r_{\rm jet}$, $\theta_{\rm jet}$, 
where $r_{\rm jet}=0$ is the position of the central 
jet source $(x,y) = {\rm (12.5\,pix , 0.5\,pix)}$, and 
polar angles $\theta_{\rm jet}$ are measured 
from N over E. The ``central'' points of the clouds
were defined by eye from the N\_Ha observation shown in 
Fig.~\ref{Haoverview}. 
The uncertainty of this procedure 
is about $\pm 1$~pixels ($\pm 3.6$~mas) for well defined (point-like) 
features (D$_{\rm SW}$), about $\pm 2$~pixels for most clouds, 
and $\pm 4$~pixels
for very elongated features like A$_{\rm N}$ or C$_{\rm SW}$ along the axis 
of elongation. These cloud positions are useful for the investigation
of radial trends or for rough relative positions 
between H$\alpha$ clouds and features seen in other observations. 

\paragraph{Temporal evolution of jet features.}
\label{Sectmotion}
Already from our two observing epochs separated by 60 days we
see for certain well defined clouds a radial motion away from
the jet source. We see also some changes in 
the brightness distributions for H$\alpha$ clouds close to
the jet source $r_{\rm jet}<0.7''$
between August and October 2014. 
The inferred outward motion is about two pixels 
for the tangential wisp C$_{\rm NE}$ and about the same for the 
point-like cloud $D_{\rm SW}$. 
Thus, the motion is of the order of 40~mas/yr 
what corresponds to a projected gas velocity of 
roughly 40~km/s. Gas motions with this speed have been
reported previously for the inner region $d<1''$ of the R Aqr jet 
\citep{Maekinen04}. The famous R Aqr jet features A and B 
located at $r_{\rm jet} \gapprox 4''$ move with an angular velocity
of $0.2''$/yr or 200~km/s significantly faster, but they
are located outside our field of view 
\citep[e.g.,][]{Hollis85,Kafatos89,Maekinen04} . 

If we compare the ZIMPOL images with the HST observations 
of \citet{Paresce94} taken 23 year earlier, then we see 
hardly a correspondence in the location of the ionized clouds 
for $r_{\rm jet}<2''$. In 1991 the NE jet was brighter than the SW jet
and the clouds in the NE were located at larger position angles
between $35^\circ$ and 60$^\circ$. 
In the SW there were only three jet clouds and
it is not clear whether these clouds just faded away or moved out 
of the central jet region ($r_{\rm jet}<2''$) since 1991. The temporal evolution of the clouds for a detailed investigation 
of gas motions and flux variations will become much clearer
from repeated measurements separated by several months to a few
years. Therefore, we postpone a discussion until we have a better temporal
coverage. 

\begin{table*}
\caption{Parameters of the jet clouds derived from the
N\_Ha (FW0) filter images from Oct. 11, 2014.} 
\label{JetTable}
\begin{tabular}{lccccccccccc}
\noalign{\smallskip\hrule\smallskip}
feature & \multispan{3}{\hfil position, size\hfil} & aperture   
               & \multispan{3}{\hfil H$\alpha$ surface brightness at $r_{\rm jet}$,$\theta_{\rm jet}$\hfil} 
               & \multispan{3}{\hfil H$\alpha$ aperture flux \hfil} \\
        & $r_{\rm jet}$ & $\theta_{\rm jet}\tablefootmark{a}$ 
               & $\ell_{\rm cl}\times w_{\rm cl}$ & $\ell_{\rm ap}\times w_{\rm ap}$ 
        & $sb_{\rm ct}({\rm N\_Ha})$ 
                          & $sb_{\rm corr}$ & ${\rm SB}_{r,\theta}({\rm H}\alpha)
                                  \tablefootmark{b}$ 
        & $f_{\rm ct}$(N\_Ha) & $ap_{\rm corr}$ & $F_{\rm cl}({\rm H}\alpha)
                                  \tablefootmark{c}$  \\
         & [$''$]  & [$^\circ$] & [pix] & [pix] 
                & ct/(s pix) &       & (b)  
                & cts/s  &      & (c) \\
\noalign{\smallskip\hrule\smallskip}
jet source  &  0     & --  &  $10\times 7$ & $30\times 24$  
            &  47.6  & $15:$  & 50700:   
            & 4550.  & 6.0 & 25.1 \\ 
\noalign{\smallskip N / NE jet\smallskip}
A$_{\rm N}$ &  0.374 & 3.2 & $70\times 15$: & $140\times 60$   
            &  2.57    & 4.3  & 785 
            & 2330    & 3.9  & 8.4 \\    
B$_{\rm NE}$ &   1.205 & 27.8  & $30\times 10$: & $50\times 30$
            &  0.030 & 6.0:  & 12.8    
            &  7.8   & 5.0   & 0.036  \\
C$_{\rm NE}$  &  1.634 & 27.4   & $60\times 10$ & $100\times 50$
            &  0.026 & 4.5:   & 8.3 
            & 25    & 4.3   & 0.099 \\    
\noalign{\smallskip SW inner jet\smallskip} 
A$_{\rm SW}$ &  0.106 & 192.7  & $12\times 10$: & $18\times 18$  
            &  0.98   & 13   & 905: 
            & 113  & 10  & 1.04 \\    
B$_{\rm SW}$  &  0.208 & 235.0  & $30\times 20$  & $52\times 32$ 
            &  1.22   & 5.5   & 477
            & 500  & 5.0  & 2.3 \\    
C$_{\rm SW}$  &  0.234 & 205.9  & $37\times 16$  & $56\times 32$
            &  3.38  & 5.5   & 1320
            & 1200 & 5.0  & 5. \\    
D$_{\rm SW}$  &  0.420 & 203.1  & $14\times 10$  & $34\times 30$
            &  0.88  & $15:$  & 938  
            &  123  & 5.5  & 0.62 \\    
\noalign{\smallskip SW outer bubbles\smallskip}
E$_{\rm SW}$  &  0.833  & 225.1 & $66\times 43$  & $100\times 100$   
            &  0.108  & 4.0   & 30.7
            & 193   & 3.8 & 0.68   \\    
F$_{\rm SW}$  &  1.225  & 217.5  &  $56\times 39$  & $100\times 100$   
            &  0.12   & 4.1   & 35.3
            & 202   & 3.8 & 0.71  \\    
G$_{\rm SW}$      &  1.405 & 232.0 & $94\times 34$:  & $200\times 120$    
            &  0.12    & 3.5: & 28.6
             & 419  & 3.5: & 1.35 \\    
H$_{\rm SW}$  & 1.853  & 217.8 & $30\times 20$ & $80\times 40$   
            & 0.029   & 5.5   & 11.3
            & 22.   & 4.5: & 0.091 \\
I$_{\rm SW}$ & 1.862  & 226.4 & $24\times 24$ & $50\times 50$ 
            &  0.016 & 5.3  & 6.0 
            &  13.   & 4.5  & 0.054  \\
J$_{\rm SW}$ & 1.935  & 234.4 & $80\times 27$ & $100\times 50$
            & 0.019  & 4.0    & 5.4 
            & 32.    & 4.3    & 0.13\\
            &       &        &               &
            &       &       &               & 
             \multispan{2}{~~~~$F_{\rm sum}$ (14 clouds)}  &  46.1 \\    
\noalign{\smallskip\hrule\smallskip}
\end{tabular}
\tablefoot{Cloud parameters are the
position $r_{\rm jet}$, $\theta_{\rm jet}$ 
of the marked cloud ``center'', length $\ell$ and 
width $w$ of the cloud and the photometric aperture, surface brightness
parameter like the background subtracted 
count rates per pixel $sb_{\rm ct}$, 
correction factor sb$_{\rm corr}$, and final surface
brightness flux SB(H$\alpha$), and cloud flux parameters for the background
subtracted count rates $f_{\rm ct}$, aperture correction factor 
$ap_{\rm corr}$ and final cloud flux $F_{\rm cl}({\rm H}\alpha)$. \\
\tablefoottext{a}{$\theta_{\rm jet}$ considers the $-2^\circ$ offset 
of the detector y-axis relative to north (Sect.~\ref{SectZimpol}),}   
\tablefoottext{b}{SB$({\rm H}\alpha)$ in units of 
$10^{-12}{\rm erg\,s}^{-1}{\rm cm}^{-2}{\rm arcsec}^{-2}$,} 
\tablefoottext{c}{$F({\rm H}\alpha)$ in units of 
$10^{-12}{\rm erg\,s}^{-1}{\rm cm}^{-2}$.}
}
\end{table*}

\subsection{Photometry for the H$\alpha$ clouds}
\label{SectHaCloudPhot}
The determination of flux parameters of individual 
H$\alpha$ features needs to take into account the 
instrument PSF and
the size of the used synthetic photometric apertures.
Therefore, the sizes of these apertures need to be 
tailored to the individual clouds and each feature 
requires its individual aperture correction. The PSF 
for ground-based, AO-assisted observations is highly variable
(Section~\ref{AOperform}) and this needs also to be taken
into account.  

Thus, the measurement of H$\alpha$ cloud fluxes is complex
and requires quite some effort. Depending on the
scientific goal of a study, one might therefore evaluate the need 
for such measurements. Knowledge of the rough 
fluxes for the individual clouds in R Aqr is certainly useful 
for estimating cloud parameters. An uncertainty of a factor
of two in the flux measurement introduces only an effect of a 
factor of $\approx 1.4$ for the determination of nebular
density from cloud emissivities. Temporal line flux variations
can be derived with a sensitivity of about 10~\%, if fluxes are
measured on a relative scale with respect to a suitable reference 
source in the image. In any case, the R Aqr commissioning ``tests''  
presented here are an ideal data set to go through this 
H$\alpha$ flux calibration exercise, because we can check our
results with the quasi-simultaneous H$\alpha$-data from HST. 
 
\paragraph{Count rates.}
For each cloud we define for the flux measurements 
synthetic round or elliptical apertures as shown 
in Figs.~\ref{innerjet}-\ref{NEwisp} with aperture sizes
$\ell_{\rm ap}$ and $w_{\rm ap}$ given in Table~\ref{JetTable}. 
In most cases the center of the aperture ellipses is close
but does not need to coincide with the astrometric point 
$r_{\rm jet}$ and $\theta_{\rm jet}$ of the cloud. The flux apertures are
optimized to include cloud extensions and to exclude contributions
from neighboring clouds, while the astrometric points pinpoint 
prominent features of the clouds.
  
The flux of an H$\alpha$ cloud $f_{\rm ct}({\rm N\_Ha})$ is calculated 
from the sum of background corrected count rates for all pixels in the aperture.
The background level is the same for all pixels and it is 
equal to the mean values derived from 
all pixels in the plotted (one pixel wide) ellipses surrounding the
synthetic aperture. This background accounts for all the 
diffuse flux from the measured cloud and the halos of all 
other H$\alpha$ emission features in the field. 
An alternative way to characterize H$\alpha$ cloud luminosities 
is via the surface brightness $sb_{\rm ct}{\rm (N\_Ha)}$ for the 
points $r_{\rm jet}$ and $\theta_{\rm jet}$ given in Table~\ref{JetTable}.

Measuring the aperture flux $f_{\rm ct}$ is accurate for strong, 
isolated, well defined clouds. For example if we compare
the counts of the bright clouds A$_{\rm N}$, B$_{\rm SW}$,
C$_{\rm SW}$, E$_{\rm SW}$ , and F$_{\rm SW}$ taken in different H$\alpha$
filters (e.g., N\_Ha and B\_Ha) or different dates (August and October), 
then the count ratios for the individual clouds scatter about 
$\sigma=10$~\% around the mean count ratio derived from 
all clouds. For faint clouds the scatter is about 20~\%. These
are good estimates of the flux measuring uncertainties which
are most likely dominated by PSF variation effects. 
 
Surface brightness measurements are also given in Table~\ref{JetTable} 
because it seems that $sb_{\rm ct}$ is a more reliable measuring quantity
for faint and diffuse clouds, but less reliable for compact 
or unresolved sources, because of
peak flux variations due to changes in the AO performance. The above obtained scatter of $\sigma\approx 10$~\% for cloud
fluxes relative to a mean value indicates that flux variation 
at this level can be recognized in repeated data, if one
emission component can be used as flux reference.

\begin{figure}
\includegraphics[trim = 2.5cm 16cm 10cm 3cm, clip, width=8.8cm]{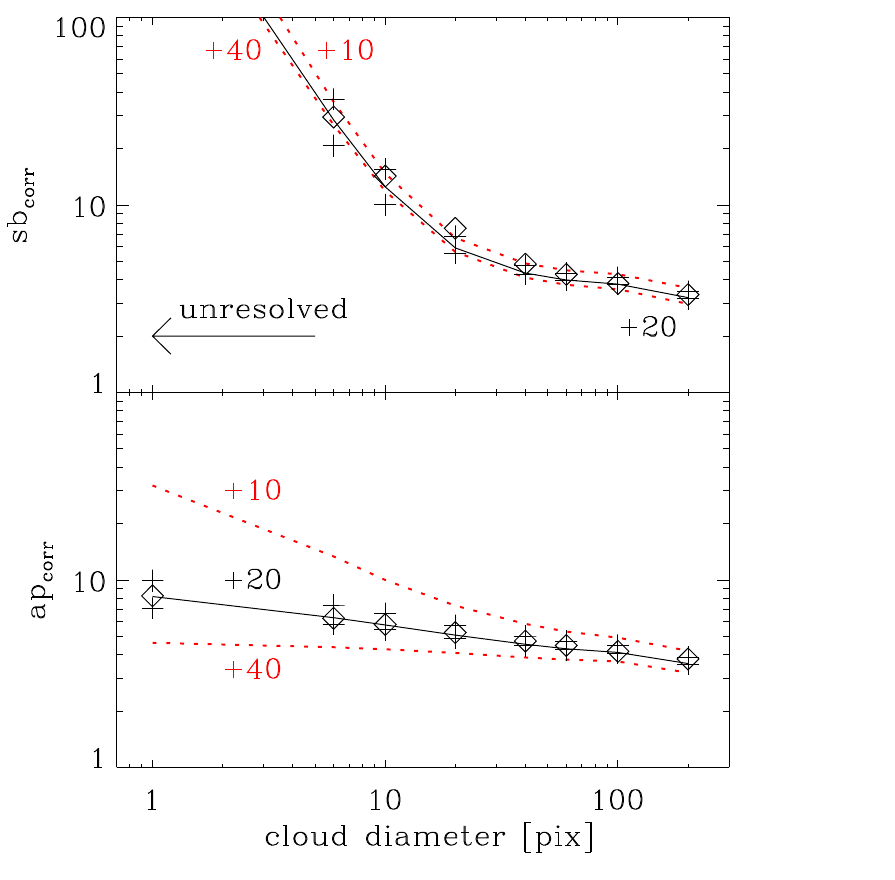}
\caption{Simulated correction factors for R Aqr H$\alpha$
surface brightness $sb_{\rm corr}$ (upper) and aperture flux $ap_{\rm corr}$ 
(lower) as
function of cloud diameter $\oslash_{\rm cl}$ [pix] for different 
apertures and AO performance. 
The full lines are for round clouds and aperture diameters 
$\oslash_{\rm ap}=\oslash_{\rm cl}+20$ and ``average'' conditions,
and crosses indicate results for good and bad atmospheric
conditions. The red dotted lines show the result for small
$\oslash_{\rm cl}+10$ or large apertures $\oslash_{\rm cl}+40$. 
Results for elliptical model clouds with the same surface area but 
a ratio $l:b=4$ are plotted with diamonds} 
\label{HaCorrFig}
\end{figure}

\paragraph{Background and aperture corrections.}
In the next step, we simulate the effect of the
extended PSF on the surface brightness and the cloud flux measurements. 
For this we simulate round and elliptical model clouds with 
constant surface brightness, the
same total flux, but different diameters. These model clouds
are convolved with the
mean PSF for the CntHa filter (OBS284\_0051-54), but also with 
the best (max) and worst (min) PSF 
(see Table~\ref{TabPSF} and Fig.~\ref{RAqrPSF}). 
Then, the net surface brightness and net cloud
fluxes are calculated for different
apertures sizes like for the R Aqr H$\alpha$ data.
The measured ratio between the initial model and PSF convolved cloud
for surface brightness or cloud flux yields then the correction 
factors, $sb_{\rm corr}=sb_{\rm mod}/sb_{\rm PSF}$ and  
$ap_{\rm corr}=ap_{\rm mod}/ap_{\rm PSF}$ respectively, as function 
of model parameters.
Figure~\ref{HaCorrFig} illustrates these model dependencies
of $sb_{\rm corr}$ and $ap_{\rm corr}$ and Table~\ref{HaCorrTab} gives 
numerical values.

The central surface brightness is obviously strongly
underestimated for measurements of unresolved and small clouds with
diameters $\oslash_{\rm cl} < 10$~pix. The correction
factor $sb_{\rm corr}$ changes much less for extended sources 
$\oslash_{\rm cl} > 20$~pix, and it depends only slightly on the 
diameter $\oslash_{\rm bck}= \oslash_{\rm cl}+ 10~{\rm pix}$, 
+20~pix (default), and +40 pix, of the ring or ellipse used for 
the background definition. In general,
larger diameters yield a slightly lower background level and
therefore a slightly higher net surface brightness requiring
a slightly smaller correction factor $sb_{\rm corr}$. 

The aperture correction $ap_{\rm corr}$ for the cloud flux depends 
quite strongly on the aperture size (see Fig.~\ref{HaCorrFig} 
and Table~\ref{HaCorrTab}). 
For point sources the correction factor is small
($ap_{\rm corr}=4.63$) for large apertures and large 
($ap_{\rm corr}=31.9$) for small apertures (see Table~\ref{HaCorrTab})
because less halo flux is included in the aperture in the latter case. 
For extended clouds the aperture size dependence diminishes. 
The impact of the PSF quality is noticeable but not very dramatic.  

\begin{table}
\caption{Simulation of correction factors for the surface brightness 
$sb_{\rm corr}$ and aperture photometry $ap_{\rm corr}$ as function of
cloud diameter $\oslash_{\rm cl}$ [pix] calculated
for the R Aqr CntHa PSF from Oct. 11, 2014.} 
\label{HaCorrTab}
\begin{tabular}{lcccccc}
\noalign{\smallskip\hrule\smallskip}
$\oslash_{\rm cl}$  &  & $sb_{\rm corr}$  & ~~ 
                           & \multispan{3}{\hfil $ap_{\rm corr}$ \hfil}  \\
       & $\oslash_{\rm bck}$=\hspace{-0.2cm}        
                  & $\oslash_{\rm cl}$+20 & $\oslash_{\rm ap}$=\hspace{-0.2cm}
                            & $\oslash_{\rm cl}$+10   
                                     & $\oslash_{\rm cl}$+20   
                                             & $\oslash_{\rm cl}$+40    \\   
\noalign{\smallskip\hrule\smallskip}
$< 1$         & & $>1000$  & &  31.9  &  8.20  &  4.63  \\
6             & & 28.5     & &  13.4  &  6.30  &  4.38  \\
10            & & 12.5     & &  10.0  &  5.76  &  4.27  \\
20            & & 5.90     & &  7.29  &  5.08  &  4.08  \\
40            & & 4.33     & &  5.83  &  4.54  &  3.86  \\
60            & & 3.98     & &  5.30  &  4.29  &  3.76  \\
100           & & 3.78     & &  4.92  &  4.11  &  3.68  \\
200           & & 3.20     & &  4.20  &  3.58  &  3.20  \\
\noalign{\smallskip\hrule\smallskip}
\end{tabular}
\tablefoot{
The diameter of the 
ring used for the sb-background determination is 
$\oslash_{\rm bck}$, and $\oslash_{\rm ap}$ is the diameter
for the circular flux aperture.}
\end{table}

The simulated correction factors can be applied to the
count rates per pixels $sb_{\rm ct}$ and the aperture count 
rates $f_{\rm ct}$ of the individual H$\alpha$ clouds. 
For this, we calculate for each cloud the average cloud diameter 
$\oslash_{\rm cl}=(\ell_{\rm cl}\cdot w_{\rm cl})^{1/2}$ and the
average aperture diameter
$\oslash_{\rm ap}=(\ell_{\rm ap}\cdot w_{\rm ap})^{1/2}$
and determine the applicable correction factors $sb_{\rm corr}$
and $ap_{\rm corr}$ from the simulation results 
in Fig.~\ref{HaCorrFig} and Table~\ref{HaCorrTab}. 
The derived correction factors are given in Table~\ref{JetTable}.

The estimated uncertainty in the determination of the 
correction factors  $sb_{\rm corr}$ and $ap_{\rm corr}$ is about $\pm 20~\%$
typically. The uncertainty is significantly larger for the surface 
brightness determination of marginally 
resolved clouds $\oslash_{\rm cl}<10$~pix because of the
strong dependence of $sb_{\rm corr}$ with $\oslash_{\rm cl}$.  

\paragraph{Surface brightness and flux for the H$\alpha$ clouds.}

The resulting H$\alpha$ cloud fluxes 
$F({\rm H}\alpha)$ and surface brightness fluxes
SB$({\rm H}\alpha)$ in Table~\ref{JetTable}
are then calculated from the measurements 
$f_{\rm ct}$, $sb_{\rm ct}$
and the derived corrections factors $ap_{\rm corr}$ 
and $sb_{\rm corr}$. These cloud fluxes are an important
measuring result of the presented observations. 

The conversion from corrected count rates into 
fluxes is given by Eq.~\ref{Eqctflux} (using
$c_{\rm zp}^\ell({\rm N\_Ha})=
8.4\cdot 10^{-16}{\rm erg}\,{\rm cm}^{-2}{\rm ct}^{-1}$,
${\rm am}\cdot k_1=0.10$, and $m_{\rm mode}=0$).  
This yields the surface brightness flux per $3.6 \times 3.6$~mas
pixel and with ${\rm pix/arcsec}^2 = 77160$ the surface 
brightness SB(H$\alpha$) per arcsec.$^2$ 

The central jet source has a flux of about 
$25 \cdot 10^{-12}{\rm erg}\,{\rm cm}^{-2}{\rm s}^{-1}$, 
which is about 55~\% of
the total flux from the central region. The sum of all H$\alpha$
clouds is 
$F({\rm H}\alpha)=46\cdot 10^{-12}{\rm erg}\,{\rm cm}^{-2}{\rm s}^{-1}$ 
in agreement with the total flux derived in Sect.~\ref{Haaperture} for
the central $3.6''\times 3.6''$ region of R Aqr. 
This is a reasonable result indicating that the 14 individual
apertures miss perhaps 10~\% or less of the H$\alpha$-emission 
in the central region, which originates from faint clouds and 
from diffuse emission. 
The polarimetric data, which will be presented in a future
paper, show that there is diffuse H$\alpha$ emission because of
dust scattering. 

The interstellar extinction towards R Aqr is small because of
the high galactic latitude of $b=-70^\circ$
and can be neglected for the interpretation of the
measured cloud fluxes and surface brightnesses. However,
circumstellar extinction is of course an issue, as
the H$\alpha$ emission regions are embedded in a
dust-rich stellar outflow. 

\subsection{The R Aqr jet in other ZIMPOL filters}

\subsubsection{[O~I] and He~I emission}

\begin{figure}
\includegraphics[trim=0.1cm 0.3cm 1.5cm 1.5cm,clip, width=8.8cm]{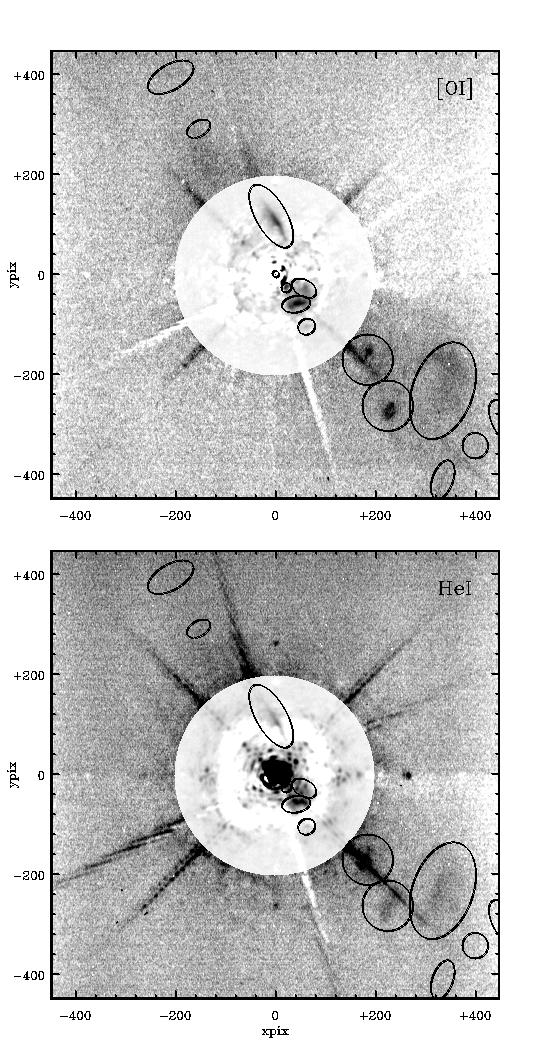}
\caption{SPHERE-ZIMPOL OI\_630- and HeI-filter observations of
the R Aqr jet. The emission of the red giant is strongly reduced
by subtracting a scaled CntHa filter image. The gray scale tables
are multiplied by a factor of ten for the inner $r_{\rm RG}<0.72''$ region.}   
\label{FigHeIOI}
\end{figure}

Besides H$\alpha$, the jet of R Aqr is also clearly detected
with SPHERE-ZIMPOL in the OI\_630-filter and the He\_I filter. 
Figure~\ref{FigHeIOI} 
shows the [O~I] and He~I observations for which the red giant 
was subtracted with scaled CntHa filter observations. 

These difference images for [O~I] and He~I are of quite 
low quality because the emission is weak and the PSF from
the CntHa observations is quite different when 
compared to the OI\_630
and He\_I data and therefore the subtraction residuals are large. 
A much better data quality for the [O~I] and He~I jet emission could
be obtained with a dedicated strategy for subtracting the 
PSF of the bright red giant. Options, which are available 
for SPHERE-ZIMPOL observations, are (i) accurate PSF-calibration
with a reference star observed with the same instrument configuration 
as R Aqr for a proper PSF subtraction, (ii) the combination of 
images taken with different field orientations to remove the
instrumental (fixed) PSF features, or (iii) angular differential imaging 
with pupil stabilized observations which would be particularly
powerful for detecting and measuring point-like emission from
a weak companion.
Simultaneous spectral differential imaging is not possible for
the OI\_630 and He\_I filters (unlike for the H$\alpha$ filters), 
because they are located in the common beam before the ZIMPOL 
beam splitter and one filter ``feeds'' both ZIMPOL arms.  

Nonetheless, the OI\_630 and He\_I filter data (Fig.~\ref{FigHeIOI})
allow a useful qualitative description. Line emission of [O\_I] is
clearly detected for the cloud components A$_{\rm N}$, the SW inner 
jet cloud C$_{\rm SW}$
and probably also B$_{\rm SW}$, and the SW outer bubbles 
E$_{\rm SW}$, F$_{\rm SW}$, G$_{\rm SW}$. 
The flux in these clouds is about ten times lower 
than the measured H$\alpha$ flux. 
An open issue with the OI\_630 filter emission is the relative
contribution of the [S\_III] 631.2~nm line, which might be responsible
for $\approx 10-50$~\% of the emission in the OI\_630 filter according
to the R Aqr spectrum shown in Fig.~\ref{LineFilters}. 

He~I emission is detected for the jet clouds A$_{\rm N}$, C$_{\rm SW}$, 
F$_{\rm SW}$, G$_{\rm SW}$ and possibly there is also some emission 
in B$_{\rm SW}$ and E$_{\rm SW}$ (affected by a diffraction spike). 
The He~I 587~nm emission is weaker by about a factor of two when compared to
[O~I] 630~nm. 

\subsubsection{Emission from the central jet source}

The central jet source is very bright in the H$\alpha$ 
emission. It is therefore of interest to search for emission
in other filters or define at least upper flux limits. 

Our simple test data, which were taken without dedicated PSF subtraction 
procedures, provide a flux contrast limit 
of  about $F_{\rm jet}/F_{\rm RG}\approx 0.05$ 
between the central jet source and the red giant. 
We identify only for the V-band filter 
observation from October 11, 2014 a source  
at the location of the central jet source. The contrast is about 
$F_{\rm jet}/F_{\rm RG}\approx 0.10$ with an estimated uncertainty of
$\pm 0.05$. This translates with red giant brightness of 
${\rm V}=11.3^m$ (Table~\ref{PhotRAqr})
into a continuum magnitude of $V_{\rm jet}\approx 14.0\,(\pm 0.7)^m$ for 
the jet source. 

In all other filters no continuum emission is detected
($F_{\rm jet}/F_{\rm RG}\leq 0.05$) from the central jet source, 
neither for the He\_I, OI\_630 line filters, nor the V\_S, 
Cnt\_Ha, TiO\_717, Cnt\_748, and Cnt\_820 intermediate band 
continuum filters. The resulting continuum magnitude
limits are $m_{\rm jet} > m_{\rm RG}({\rm Tab.~\ref{PhotRAqr}}) + 3.25^m$ 
when taking the red giant magnitudes from Table~\ref{PhotRAqr}.  

The detection in the V-band yields, with the distance modulus 
of $m-M=6.7^m$ for R Aqr, an absolute magnitude of about 
$M_{\rm V}=7.3^m$. This is brighter than expected for a hot
white dwarf on the cooling track and more compatible with
a star of the heterogeneous class of O or B subdwarfs,
which are mass accreting and ``active'' in interacting binary 
systems \citep[e.g.,][]{Heber09}. According to historical
Korean nova records, the hot component could
have had a nova like outburst in 1073 and 1074 \citep{Yang05}. 
Thus, the hot star in R Aqr could be an
accreting compact object on its evolutionary track from
a symbiotic nova outburst  
towards a cold and less luminous white dwarf state \citep[see][]{Murset94}.  

The observations in the He\_I and OI\_630 
line filters originate from August 2014, when the red giant was about 
one magnitude brighter than in October 2014. We can define
line flux limits relative to H$\alpha$, but they are with
$F({\rm He~I})/F({\rm H}\alpha)<0.15$ and 
$F([{\rm O~I}])/F({\rm H}\alpha)<0.15$ not sensitive. These
limits are compatible with the expected line emission from an
ionized gas nebula. 

With ZIMPOL / SPHERE high contrast observations, using, for example, pupil stabilized
angular differential imaging, it should be possible to
reach a contrast limit of about 5~mag between the red giant and the
jet source for the current angular separation, or line flux ratios 
$<0.05$ (relative to H$\alpha$). Thus, a better characterization 
of the jet source is certainly possible if dedicated observations 
are carried out during the luminosity minimum of R Aqr.

\section{HST line filter observations}

\subsection{HST / WFC3 Data}

\object{R Aqr} was also observed in 2013 and 2014 with the Ultraviolet-Visible
(UVIS) channel of the HST Wide Field Camera 3 (WFC3). Long and short exposures 
were taken in the four line filters
F502N, F631N, F656N, and F658N targeting the nebular emission lines 
[O~III] 500.7~nm, [O~I] 630.0~nm, H$\alpha$ 656.3~nm, and [N~II] 658.3~nm. 
Exposures with ``long'' integration times of either 1086~s or 2085~s 
are saturated in the center, but they show  
the extended jet and nebulosity at $r>1''$ with very high sensitivity 
as shown for H$\alpha$ in Fig.~\ref{HSTZimpol}a. The ``short'' 
exposures with $t_{\rm exp}$ between 15~s and 70~s are not saturated 
and such data were taken in all four
filters on October~18, 2014 only seven days after our 
observation from October~11. Therefore, the HST data
complement in an ideal way our SPHERE-ZIMPOL observations with
additional, quasi-simultaneous line measurements 
for [O~III] and [N~II], and higher
sensitivity H$\alpha$ and [O~I] images. The HST provides high quality
flux calibrations which is particularly important for a cross check of 
our ZIMPOL H$\alpha$ flux measurement and calibration procedure.   

Table \ref{HSTdata} lists the parameters for the unsaturated 
HST data selected for this work. The spatial resolution of HST-UVIS
is about 78~mas (two pixels). We consider only the 
central R Aqr region of about $3.5''\times 3.5''$ or $90\times 90$ pixels
which is shown in Fig.~\ref{HSTfilters} for H$\alpha$, [O~III], and
[O~I]. Of course, the entire HST field of view contains a lot of important
information about the outer jet and nebula of R Aqr, but this is
beyond the scope of this paper.

\begin{table}
\caption{HST WFC3 observations of R Aqr from Oct. 18, 2014 used
for this work.} 
\label{HSTdata}
\begin{tabular}{lcccc}
\noalign{\hrule\smallskip}
filter  &  F502N  & F631N  & F656N  & F658N \\      
\noalign{\smallskip\hrule\smallskip}  
ident ``ic9k0...''    & 6vnq  &    7vwq   & 6010 &   7010 \\
$t_{\rm exp}$      & 50~s       &  24~s       & 70~s      & 70~s \\
$\lambda_c$ [nm]  & 501.0      & 630.4      & 656.3     & 658.5  \\
$\Delta\lambda$ [nm] & 6.5     & 5.8        & 1.8      & 2.8 \\
Transm. $T$      & 0.230      & 0.230       & 0.228    & 0.245   \\
target line      & [O~III]    & [O~I]       & H$\alpha$  & [N~II]  \\
~~~~other line   &  -       & [S~III]     & [N~II]     & H$\alpha$ \\
\noalign{\smallskip\hrule\smallskip}
\end{tabular}
\tablefoot{Observations are identified with filter names in
the column heading, while the rows give frame identifications 
(all have the prefix ``ic9k0...''), 
exposure times, and the filter parameters central wavelengths 
$\lambda_{\rm c}$, filter widths $\Delta\lambda$ , and the total 
system transmissions at the wavelength of the targeted lines.}
\end{table} 
 
For our analysis we started with the pipeline processed images
as they can be retrieved from the HST-MAST archive. In the 
innermost R Aqr region, the same H$\alpha$ cloud structures can 
be recognized as in the ZIMPOL H$\alpha$ images.  
Because of the lower resolution of the HST data, the clouds A$_{\rm SW}$ 
and D$_{\rm SW}$ cannot be resolved and the clouds B$_{\rm SW}$ and 
C$_{\rm SW}$ are merged into one single feature, which we call (B+C)$_{\rm SW}$. 
  
\subsection{Line flux measurements for the HST data}

\begin{figure*}
\includegraphics[trim=0.1cm 0.2cm 0.1cm 0.2cm,clip,width=18cm]{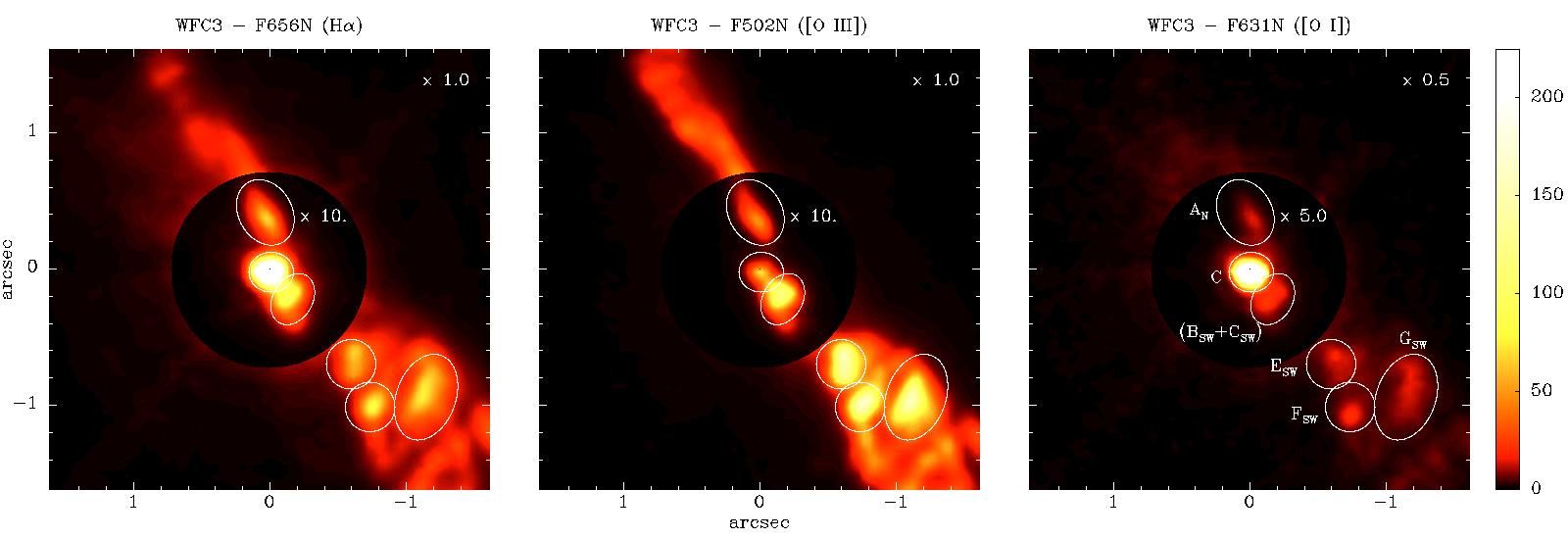}
\caption{HST images of the central region of R Aqr taken in the
H$\alpha$, [O~III], and [O~I] filters. The color scale of the inner region 
is ten times larger than the outer region and factors for the application of
the color scale are indicated. The cloud features are identified in the
O~I panel.}
\label{HSTfilters}
\end{figure*}

The HST observations of R Aqr taken with the 
H$\alpha$, [O~III], [N~II], and [O~I] line filters 
show quite significant differences for the innermost clouds
(see Fig.~\ref{HSTfilters}). 
The strongest emission in the red filters F631N, F656N, 
and F658N originates from the central, unresolved stellar binary. 
Interestingly, the stellar binary is rather weak in the [O~III] line, 
weaker than the clouds A$_{\rm N}$ and (B+C)$_{\rm SW}$, 
indicating that the [O~III]/H$\alpha$ ratio
varies strongly between stellar binary and jet clouds. 
The jet clouds are much fainter in [O~I] than in H$\alpha$ confirming
the result from SPHERE / ZIMPOL, and also the [N~II] line is
much weaker than H$\alpha$ or [O~III] (see Table~\ref{HSTLinetable}).

For the line flux measurements and the comparison with the ZIMPOL
data, we have selected six well defined emission features in the HST data
which are indicated with elliptical apertures in Fig.~\ref{HSTfilters}. 
These are the stellar binary SB, and the clouds A$_{\rm N}$, (B+C)$_{\rm SW}$,
E$_{\rm SW}$, F$_{\rm SW}$, and G$_{\rm SW}$ as defined in the ZIMPOL
H$\alpha$ observations.

The apertures for the flux measurements were chosen to be similar to the 
used ZIMPOL cloud apertures, if possible.
For this, each $39.6 \times 39.6$ mas pixel of the HST image was expanded 
into $11 \times 11$ pixels with a size of $3.6 \times 3.6$ mas 
with a flux conserving interpolation. The expanded images, which
are actually displayed in Fig.~\ref{HSTfilters} together
with the flux apertures, have the pixel scale of the ZIMPOL data 
so that the same type of measuring routines could be used. 

The resulting line fluxes are given in Table~\ref{HSTLinetable}.
Fluxes were calculated from the summed count rates
ct/s (called e$^-$/s in the HST jargon) in a given cloud aperture using
the transformation  
\begin{displaymath}
F_{\rm cl}={\rm ct/s}\cdot\, {1\over A\cdot T}\,{hc\over \lambda}\cdot c_{\rm ap} \,, 
\end{displaymath}
where $A=45239~{\rm cm}^2$ is the effective HST primary mirror area, $T$
the HST-WFC3 system transmission, 
$hc/\lambda$ the line photon energy,
and $c_{\rm ap}$ the aperture correction for a given cloud. 
For each filter $T$ is listed in Table~\ref{HSTdata} and as derived
from the transmission curves given in the WFC3 handbook.  

The aperture corrections $c_{\rm ap}$ are estimated as for the ZIMPOL data 
(see Sect.~\ref{SectHaCloudPhot}) by calculating 
the halo (or background) corrected energy within given apertures
for simple models of extended clouds, convolved with the HST point 
spread function. The derived $c_{\rm ap}$-values 
listed in Table~\ref{HSTLinetable} for the
different clouds are quite large, 
because small apertures must be used to avoid an overlap with 
neighboring clouds. The $c_{\rm ap}$ correction factors introduce
an estimated uncertainty of about $\pm 10-20$~\% 
and this dominates
the errors in the line fluxes given in Table~\ref{HSTLinetable}.

Line flux ratios, for example, $F({\rm [O~III]})/F({\rm H}\alpha)$, 
from HST are expected to be very accurate ($\pm 10~\%$), 
because systematic line measuring errors, for example, due to the $c_{\rm ap}$
correction factors, are strongly reduced for line ratios. 
However, one should also
note that the emission regions for different lines do not coincide
always because of the complex substructure of a cloud. For example
the clouds F$_{\rm SW}$ and G$_{\rm SW}$ show different structures 
for [O~III] and [O~I] (see Fig.~\ref{HSTfilters}). 

\paragraph{HST line fluxes for the central jet source.}  
The mira variable and the central jet source are not resolved
in the HST data. Therefore we need to correct for the determination
of the line flux for the central jet source the contribution of
the red giant to the flux in the SB-aperture. 

Table~\ref{HSTLinetable} gives a value
``SB total'' which would corresponds to the line flux from the
region covered by the central stellar binary (SB) aperture assuming that 
all photons emitted from this region are line photons. This is
certainly not true, because the contribution from the red giant
continuum is at least for the red line filters substantial. 
The following line gives an estimate for the
relative contribution of the red giant to the flux in the SB aperture
as discussed in the following paragraphs and the third line
is then the resulting line flux estimate for the central jet
source, if the contribution of the red giant is taken into account.  

For H$\alpha$ and [O~I], the contribution of the
red giant can be estimated from the ZIMPOL observations from which we
measured for October~11 a relative photon ratio of 1.1$\pm 0.1$ between
the jet source and the red giant in the N\_Ha filter. The H$\alpha$ 
filter F656N of HST has twice the width of the ZIMPOL N\_Ha filter
and  therefore the contribution
of the red giant is roughly 65~\% of the total flux
in the SB-aperture in the H$\alpha$ HST image. 

The [O~I] line was not detected for the central jet source with 
the ZIMPOL on August~12. The estimated contrast limit 
is [O~I]/RG~$<0.05$, which turns into a contrast
limit of about $<0.1$ for the October~18 HST epoch when the red giant was
fainter. The [O~I] filters in ZIMPOL and HST have roughly 
the same widths and one can assume as a conservative limit that less than 
20~\% of ``SB total'' value for the F603N filter for the SB aperture 
originates from nebular line emission. 

For [N~II] a similar red giant
continuum flux as for the H$\alpha$ filter can be assumed. This
yields an expected photon count rate for the red giant in the F658N filter 
which is slightly above the measured count rates. Thus,
the [N~II] emission line flux is low and a  
conservative line flux limit is indicated in Table~\ref{HSTLinetable}.

The contribution of the red giant in the F501N filter is unclear and
difficult to estimate. Therefore, we use as a conservative upper 
limit for the [O~III] line flux the ``SB total'' line flux.

\begin{table}
\caption{HST emission line fluxes for R Aqr emission clouds derived
from the H$\alpha$, [O~III], [N~II] and [O~I] filter observations.}
\label{HSTLinetable}
\begin{tabular}{lccccc}
\noalign{\smallskip\hrule\smallskip}
cloud        & $c_{\rm ap}$ & H$\alpha$  & [O~III]  & [N~II]  & [O~I]  \\
\noalign{\smallskip\hrule\smallskip}
A$_{\rm N}$     &   1.4     &  5.5      & 4.8    & 0.51    & 0.60   \\ 
(B+C)$_{\rm SW}$ &   1.9     &  5.1      & 10.0   & 0.25    & $<0.5$ \\ 
E$_{\rm SW}$    &   1.8     &  0.41     & 1.8    & 0.030   & $<0.1$ \\
F$_{\rm SW}$    &   1.8     &  0.54     & 1.5    & 0.11    & 0.046  \\ 
G$_{\rm SW}$    &   1.4     &  0.89     & 2.2    & 0.35    & 0.42   \\
\noalign{\smallskip central jet source \smallskip}
SB total     &   2.0     &  $34$    & $2.6$  & $16$   & $15$  \\
rel. RG cont.&          &  $\approx 65~\%$
                                     & unclear & $>80~\%$ & $>80~\%$ \\   
C (jet source) &       &  $\approx 12$
                                     & $<2.6$  & $<3$    & $<3$ \\
\noalign{\smallskip\hrule\smallskip}

\end{tabular}
\tablefoot{Line fluxes are given
in units of $10^{-12}{\rm erg}\,{\rm cm}^{-2}{\rm s}^{-1}$ and 
$c_{\rm ap}$ are the applied flux aperture correction factors
for the individual cloud. The last three lines
give the line flux ``SB total'' assuming all flux is 
line emission, an estimate for the
relative red giant contribution, and the resulting line
flux estimate for the central jet source ``C''.}
\end{table}

\paragraph{H$\alpha$ flux comparison between HST and ZIMPOL.}
We can now compare the derived H$\alpha$ cloud fluxes derived in 
Table~\ref{HSTLinetable} for the HST observation with 
Table~\ref{JetTable} derived from the ZIMPOL observations.
The obtained mean H$\alpha$ flux ratio is 
$F_{\rm HST}/F_{\rm ZIMPOL}=0.67\pm 0.05$ for the five clouds $A_{\rm N}$,
$(B+C)_{\rm SW}$, $E_{\rm SW}$, $F_{\rm SW}$, $G_{\rm SW}$. 
This is quite a significant difference.  
However, the relative scatter of $\sigma/{\rm mean} \approx 7.5~\%$ 
in the derived flux ratios $F_{\rm HST}/F_{\rm ZIMPOL}$ for the five clouds 
is very small. Thus, we can conclude that the H$\alpha$ flux
ratios between individual clouds, like $F(E_{\rm SW})/F(A_{\rm N})$ ,
agree very well between HST and ZIMPOL-SPHERE.   

The large overall H$\alpha$ flux difference is hard to explain.  
The main uncertainties are the aperture correction factors derived
for the ZIMPOL measurements and the preliminary flux zero-points 
calibration of ZIMPOL, which is not well established yet.
More investigation is required to clarify this issue.

\section{Physical parameters for the H$\alpha$ clouds}

\subsection{Temperatures and densities for the jet clouds}

In this section we derive nebular densities $N_e$ and temperatures $T_e$ 
for the jet clouds from the measured
ZIMPOL H$\alpha$ line emissivities $\epsilon({\rm H}\alpha)$ 
and the HST [O~III]/H$\alpha$ line ratios. The
H$\alpha$ emissivity provides a good measure of the nebular
density, and the combination with the [O~III]/H$\alpha$ ratio yields
the nebular temperature.   
\smallskip

\noindent
For the theoretical line emissivities 
$\epsilon({\rm H}\alpha),$ we assume that the H$\alpha$ line 
is produced mainly by case B recombination \citep{Osterbrock06,Hummer87}. 
Case B assumes that the lower H\,I Lyman lines are optically thick for 
the jet clouds in R Aqr which corresponds to H\,I column densities of
roughly 
$\gapprox 10^{14}\,{\rm cm}^{-2}$. Case B conditions might not be
fulfilled because the emission clouds in R Aqr are small 
($r\approx 10^{14}$~cm), and there is significant line broadening 
due to gas motions $v > 100$~km/s so that Lyman line photons may escape.
Thus, case A (optical thin Lyman lines) might apply and the
corresponding recombination emissivities would be lower 
by a factor of about 0.67. 

On the other side, there could be an enhancement of 
the H$\alpha$ line emissivities by collisions 
from the ground state or the metastable state H\,I $^2$S. 
Collisions from the ground state are important for high
nebular temperatures ($T> 10000$~K) near a shock front 
\citep[see, e.g.,][]{Raymond79,Hartigan87,Raga15a} but this effect
can be strongly suppressed if X-rays from the shock
reduce strongly the density of H$^0$ by photo-ionization 
in that region. In shock models and observations of Herbig-Haro 
objects, most of the H$\alpha$ emission originates from recombination 
in the cold ($T_e\lapprox 10000$~K) post-shock cooling region 
\citep[e.g.,][]{Raymond79,Raga15b}. 
In R Aqr the nebular densities in the innermost jet region are
of the order $10^6\,{\rm cm}^{-3}$, which is at least two orders of
magnitude higher when compared to typical Herbig-Haro objects.
For R Aqr several studies on the jet emission exist 
\citep{Burgarella92,Kellogg07,Nichols09} for line emitting clouds 
located at large separations $> 1000$~AU, several times further out 
than the clouds studied in this work. There the conditions are 
comparable to Herbig-Haro objects
with $T_e\approx 10^4$~K, $N_e\approx 10^4{\rm cm}^{-3}$ will the X-ray 
emitting material has parameters of $T_e\approx 10^6$~K and 
$N_e\approx 10^2{\rm cm}^{-3}$.

The line emission of the central nebula of R Aqr was
investigated by \citet{Contini03} with a (plane parallel) shock model 
describing a scenario where a fast (preionized) wind from the 
hot component with $v=110$ -- $125$~km/s and a high pre-shock density of 
$6\cdot 10^4{\rm cm}^{-3}$ collides with the wind of the red giant.
They obtain for the main emission region (post-shock cooling region) 
a temperature of $T_e\approx 10^4$~K and a density of 
$\geq 6\cdot 10^5{\rm cm}^{-3}$. This indicates that the resulting
H$\alpha$-emission in the jet clouds imaged by us 
originates mainly from recombination. 

For these reason it seems
reasonable to adopt the case B recombination emissivities as
first approximations for the ``theoretical'' H$\alpha$ emissivities for our
study of R Aqr, but considering the complexity of the H$\alpha$
line formation we admit an uncertainty of a factor of two in 
$\epsilon_{\rm th}({\rm H}\alpha)$. The impact on the determination of 
the nebular density is then about 
an uncertainty of a factor of 1.4.
\smallskip

\noindent
The adopted ``theoretical'' H$\alpha$ emissivity is 
\begin{equation}
\epsilon_{\rm th}({\rm H}\alpha) = \alpha_B({\rm H}\alpha,T_e)\, h\nu\, N_pN_e\,,
\label{emissHa}
\end{equation}
%
where $\alpha_B({\rm H}\alpha,T_e)\approx 
7.86\cdot 10^{-14}\,{\rm cm}^3{\rm s}^{-1}
\cdot (T_e/10000~{\rm K})^{-0.9}$  
are the case B recombination coefficients for the H$\alpha$ line, 
$T_e$ the electron temperature, $h$ the
Planck constant, $\nu$ the photon frequency, and $N_p,N_e$ the
proton and electron densities \citep[see][]{Osterbrock06}. We approximate
$N_p\approx N_e$, assuming that the jet clouds are
strongly ionized.

Averaged H$\alpha$ emissivities $\langle \epsilon({\rm H}\alpha)\rangle$
for a given cloud can be derived from the observed line flux 
$F({\rm H}\alpha)$, the distance $d$ to R Aqr, and the cloud 
volume $V_{\rm cl}$ estimated from the measured cloud size
\begin{equation}
\langle \epsilon({\rm H}\alpha) \rangle \approx {4\pi d^2 F({\rm H}\alpha)
\over V_{\rm cl}}\,.
\end{equation}
The cloud volume is calculated according to 
$V_{\rm cl} = \pi\,(\ell_{\rm cl}\cdot w_{\rm cl})^{3/2}/6$, where cloud 
lengths $\ell_{\rm cl}$ and widths $w_{\rm cl}$
are taken from Table~\ref{JetTable}. This assumes a cloud diameter 
along the line of sight which is equivalent to 
$s_{\rm cl}\approx (\ell_{\rm cl}\cdot w_{\rm cl})^{1/2}$. Thus, we define
$\oslash_{\rm cl}=(\ell_{\rm cl}\cdot w_{\rm cl})^{1/2}$ as the equivalent 
diameter for a 
spherical cloud with the same volume.
This approximation would introduce a bias if the H$\alpha$-clouds in R Aqr 
have typically filamentary or strongly flattened structures.

An alternative determination of $\langle \epsilon({\rm H}\alpha)\rangle$
can be made from the measured surface brightness SB(H$\alpha$) 
using the relation
\begin{equation}
\langle \epsilon({\rm H}\alpha) \rangle 
\approx {\Omega\over 4\pi} {{\rm SB}({\rm H}\alpha)
\over s_{\rm cl}}\,.
\end{equation}
For the SB-values from Table~\ref{JetTable}, which are given per arcsec$^2$, 
one must use $\Omega/4\pi = 1.87 \cdot 10^{-12}$. 
As above, we use $s_{\rm cl}\approx \oslash_{\rm cl}$
for the line of sight diameter of a given cloud. 

Table~\ref{Nedens} lists the resulting emissivities  
$\langle \epsilon({\rm H}\alpha)\rangle$ using
the cloud flux $F_{\rm cl}$ or surface brightness SB$_{\rm cl}$
data from Table~\ref{JetTable}. Both methods yield similar results
$\epsilon_F/\epsilon_{\rm SB}\approx 1.4$ with a scatter of 
0.15 dex. We select for well defined bright clouds the 
emissivities derived from $F_{\rm cl}$ and for diffuse and faint clouds the
values from SB$_{\rm cl}$, because these values seem to be
less affected by background determination or
aperture definition uncertainties. The obtained emissivities indicate
nebular densities in the range $N_e\approx 10^5-10^8\,{\rm cm}^{-3}$
for the clouds $r_{\rm jet}<2''$ ($< 400$~AU) of the R Aqr jet. 

The nebular temperature can be constrained by the [O~III]/H$\alpha$
line ratio derived from the HST data. The [O~III] 500.6 nm line 
is a collisionally excited line and its emissivity is described by 
\begin{equation}
\epsilon({\rm [O~III]}) = N_u\, h\nu\, A_{ul} 
     =  N_{{\rm O}^{+2}}\, h\nu\, A_{ul} \, {N_e q_{lu} \over A_{ul}+N_e q_{ul}}\,, 
\label{emissO3}
\end{equation}
where $N_u$ is the density of O$^{+2}$-atoms in the upper state ($^1$D) 
of the [OIII] 500.7~nm transition, $h\nu=hc/\lambda$ the corresponding 
photon energy, and $A_{ul}$ the transition rate of the line. The
level population $N_u$ is defined by the balance of collisional excitations
$N_e q_{lu}$ from the lower states (ground level term), and the collisional
de-excitations $N_e q_{ul}$ and radiative decays $A_{\rm ul}$
\citep{Osterbrock06}. The [O~III] emissivity is very sensitive to the 
temperature because of the exponential temperature term in the 
collisional excitation $q_{lu}\propto {\rm exp}(-h\nu/kT_e)$. 

For low density gas the emissivity of collisionally excited lines
is $\epsilon\propto N_{{\rm O}^{+2}}N_e$ or proportional to the density
squared. However, the central clouds in R Aqr 
are very dense, above the critical density of the [OIII] 500.7 nm 
transition, and therefore collisional de-excitations are 
very important for $A_{ul} \ll  N_e q_{ul}$. In this high density regime, the
emissivity is $\epsilon({\rm [O~III]})\propto N_{{\rm O}^{+2}}{\rm exp}(-h\nu/kT_e)$ or 
proportional to the density of the O$^{+2}$ atoms. 
\smallskip

\noindent
We can relate $N_{{\rm O}^{+2}}$ to the electron density $N_e$ with the
approximation
\begin{displaymath}
N_{{\rm O}^{+2}}={N_{{\rm O}^{+2}}\over N_{\rm O}}{N_{\rm O}\over N_{\rm H}}
       {N_{\rm H}\over N_e} \, N_e \approx 5\cdot 10^{-4}\, N_e\,.
\end{displaymath} 
This assumes for the clouds that hydrogen is strongly 
ionized $(N_{\rm H}/N_e\approx 1)$, oxygen is predominantly in 
the form O$^{+2}$ ($N_{{\rm O}^{+2}}/ N_{\rm O}\approx 1$),
and the relative abundance of oxygen to hydrogen is solar 
$N_{\rm O}/N_{\rm H}=5\cdot 10^{-4}$. All clouds for which
we have measured line fluxes from the HST data (Table~\ref{HSTLinetable})
show much stronger [O~III] than [O~I] or [N~II] lines, 
what supports the assumption of a high degree of ionization. 

\begin{figure}
\includegraphics[trim=2cm 12.5cm 1.8cm 3cm,clip,width=8.8cm]{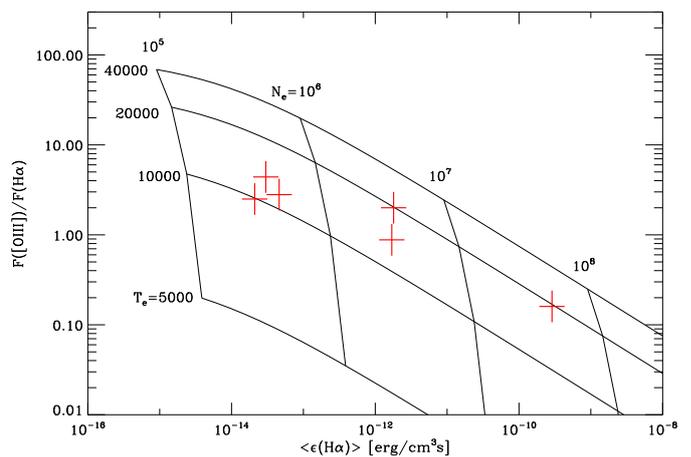}
\caption{Diagnostic diagram based on H$\alpha$ line emissivities
$\epsilon({\rm H}\alpha)$ and [O~III]/H$\alpha$ line ratios assuming
an abundance ratio $N_{{\rm O}^+}/N_{{\rm H}^+}=5\cdot 10^{-4}$. The crosses
give the values for the H$\alpha$ clouds with [O~III]/H$\alpha$-ratios
from HST.}
\label{FigHaO3}
\end{figure}

A diagnostic diagram for the determination of nebular densities
and temperatures from the ZIMPOL H$\alpha$ emissivities
$\langle \epsilon({\rm H}\alpha)\rangle$ 
and the $F([{\rm O~III}])/F({\rm H}\alpha)$ line ratios from HST
is plotted in Fig.~\ref{FigHaO3} for the clouds with HST line 
flux measurements. 
The diagram is based on simple theoretical
line emissivities described above 
(Eqs.~\ref{emissHa} and \ref{emissO3}) using the 
atomic data compiled in \citet{Osterbrock06}, and adopting 
a solar abundance ${N_{{\rm O}^{+2}}/N_e}=5\cdot 10^{-4}$. 

The considered clouds show a range of
about a factor of 100 in density from about 
$N_e\approx 3\cdot 10^5{\rm cm}^{-3}$ to $3\cdot 10^7{\rm cm}^{-3}$,
while the derived temperatures are between about $T_e=10'000$~K and
$20'000$~K. Because of the collisional de-excitation of the
nebular [O~III]-line, the $F([{\rm O~III}])/F({\rm H}\alpha)$ line ratio
decreases rapidly for increasing density. A high density
can therefore naturally explain the weak [OIII]-emission from the 
central jet source.

We can assume, based on Fig.~\ref{FigHaO3}, that all the 
H$\alpha$-clouds in Table~\ref{Nedens} have a temperature
of roughly $T_e\approx 15000\pm 5000$~K. 
Adopting this value for clouds without [O~III]/H$\alpha$ 
ratio yields densities for all these clouds.

\begin{table*}
\caption{Parameters for the H$\alpha$ clouds in the jet of R Aqr for
Oct.~11, 2014.}
\label{Nedens}
\begin{tabular}{lccccccccc}
\noalign{\smallskip\hrule\smallskip}
cloud        & $r_{\rm RG}$ & $\oslash_{\rm cl}$ & method & $\epsilon(H\alpha)$ 
             & $F$-ratio   & $T_e$ & $N_e$  & $M_{\rm cl}$ \\ 
             & $10^{13}$~[cm] & $10^{13}$~[cm] & $F/{\rm SB}$ & $10^{-12}{\rm erg/cm}^{-3} $ 
             & [O~III]/H$\alpha$ & [K]      & $10^6{\rm cm}^{-3}$   & [10$^{-9}$M$_\odot$] \\
\noalign{\smallskip\hrule\smallskip}
C (center)   & 14.3  & 9.8    & $F_{\rm cl}$        &  288    
             & $<0.2$  & 20\,000 &   44     &    24. \\
\noalign{\smallskip N / NE jet\smallskip}   
A$_{\rm N}$   & 121   & 38     & $F_{\rm cl}$        &  1.7
             & 0.87   & 13\,000 &   2.9     &   91. \\       
B$_{\rm NE}$  & 385   & 20     & ${\rm SB}_{\rm cl}$ &  0.034
             &       & \tablefootmark{a}    &   0.43     &  2.1 \\     
C$_{\rm NE}$  & 525   & 29     & ${\rm SB}_{\rm cl}$ &  0.015
             &       & \tablefootmark{a}    &   0.29     &  4.0  \\     
\noalign{\smallskip SW inner jet\smallskip}
A$_{\rm SW}$  & 41.5  & 13     & $F_{\rm cl}$        &  5.3 
             &       & \tablefootmark{a}    &   5.4     &   6.6 \\      
B$_{\rm SW}$  & 82.1  & 29     & $F_{\rm cl}$        &  1.05
             & 2.0$^a$ & 20\,000 &   2.7     &   36. \\      
C$_{\rm SW}$  & 85.0  & 29     & $F_{\rm cl}$        &  2.6
             & 2.0$^a$ & 20\,000 &   4.1    &    56. \\       
D$_{\rm SW}$  & 144   & 14     & $F_{\rm cl}$        &  2.5
             &       & \tablefootmark{a}    &   3.8     &   5.8  \\       
\noalign{\smallskip SW outer bubbles\smallskip}
E$_{\rm SW}$  & 283   & 63     & $F_{\rm cl}$        &  0.030
             & 4.4   & 13\,000 &   0.39     &   54.  \\     
F$_{\rm SW}$  & 410   & 55     & $F_{\rm cl}$        &  0.046
             & 2.8   & 11\,000 &   0.46     &   43.  \\     
G$_{\rm SW}$  & 471   & 66     & $F_{\rm cl}$        &  0.021
             & 2.5   & 10\,000 &   0.46     &   77.  \\     
H$_{\rm SW}$ & 614    & 29     & ${\rm SB}_{\rm cl}$ &  0.026
             &       & \tablefootmark{a}     &   0.34     &   4.7  \\     
I$_{\rm SW}$ & 619    & 29     & ${\rm SB}_{\rm cl}$ &  0.011
             &       & \tablefootmark{a}     &   0.25     &   3.2  \\     
J$_{\rm SW}$ & 644   & 55     & ${\rm SB}_{\rm cl}$ &  0.0053
             &       & \tablefootmark{a}     &   0.17     &   16.  \\    
\noalign{\smallskip}
uncert.     &       & $\pm 0.05$~dex &        & $\pm 0.25$~dex
            & $\pm 0.05$~dex &        & $\pm 0.30$~dex 
                                                   & $\pm 0.45$~dex \\
\noalign{\smallskip\hrule\smallskip}
\end{tabular}

\tablefoot{The given cloud parameters are: distance from the
red giant $r_{\rm RG}$, mean cloud diameter $\oslash_{\rm cl}$, 
H$\alpha$-emissivity $\epsilon({\rm H}\alpha)$,  
$F([{\rm O~III}])/F({\rm H}(\alpha)$-line ratios, electron
temperature $T_e$ and density $N_e$ and the cloud mass $M_{\rm cl}$. The last 
line indicates typical uncertainty factors for the cloud parameters.
\\
\tablefoottext{a}{adopted temperature $T_e=15\,000\pm 5000$~K}
} 
\end{table*}

\subsection{Cloud density versus cloud distance}

The H$\alpha$ clouds in the inner jet region of R Aqr show 
a systematic decrease in surface brightness and cloud flux with
distance from the central jet source (e.g., Fig.~\ref{Haoverview}). 
This points to a relation between distance and cloud density,
because the emissivity is $\epsilon({\rm H})\propto N_e^2$.

\begin{figure}
\includegraphics[trim=1.5cm 12.5cm 2cm 3cm,clip,width=8.5cm]{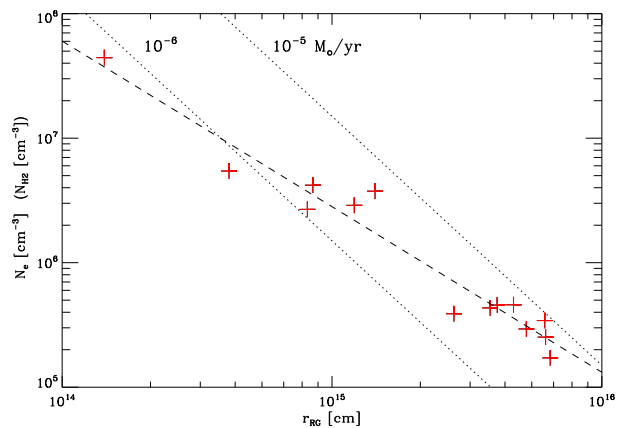}
\caption{H$\alpha$ cloud densities $N_e$ as function of
distance to the red giant $r_{\rm RG}$. The dashed line is the power law
fit $N_e\propto r_{\rm RG}^{-1.3}$ to the derived values. The dotted lines
illustrate the expected molecular hydrogen density $N_{{\rm H}_2}(r_{\rm RG})$ in
the red giant wind for a wind velocity of $v_\infty=10$~km/s and 
mass loss rates $\dot{M}=10^{-5}$ and $10^{-6}{\rm M}_\odot$/yr.}
\label{HaRNe}
\end{figure}

Figure~\ref{HaRNe} shows the cloud parameters from Table~\ref{Nedens}
in the distance - density plot and the points are well fitted
with a power law $N_e = N_{14}\, r_{\rm RG}^{-1.3}$ with the
normalization density $N_{14}=6\cdot 10^7 {\rm cm}^{-3}$
at the distance $r_{\rm RG}=10^{14}{\rm cm}$.
Essentially the same correlation is found if the point for the
central jet source at $r_{\rm RG}= 1.49\cdot 10^{14}{\rm cm}$ is
not considered for the least square fit, or if the density is plotted
versus jet source distance $r_{\rm jet}$ instead of $r_{\rm RG}$ 
(omitting the jet source).   

The tight correlation between density $N_e$ and cloud distance 
from the red giant $r_{\rm RG}$ is indicative of a kind of 
pressure equilibrium between the neutral gas of the red giant wind
and the ionized H$\alpha$ clouds in the jet. The molecular hydrogen
density $N_{{\rm H}_2}$ in an isothermal spherical wind from the
red giant can be estimated from the formula 
$N_{{\rm H}_2}(r)=\dot{M}/(4\pi\,\mu\cdot m_{\rm H}\,v_\infty r^2)$,
where we adopt $\mu=2$ for the mean atomic weight of the particles, and 
$m_{\rm H}$ is the mass of the hydrogen atom.
Figure~\ref{HaRNe} includes the lines for the expected $N_{{\rm H}_2}(r)$
density dependence, if the mira has a spherical mass loss 
of $\dot{M}_{\rm RG}=10^{-5}$ or $10^{-6}\,{\rm M}_\odot/{\rm yr}$ and a wind 
velocity of $v_\infty=10~{\rm km/s}$. Unfortunately, the mass loss 
rate for the mira variable in R Aqr is not well known and literature 
values range from about $\dot{M}_{\rm RG}=10^{-7}\,{\rm M}_\odot/{\rm yr}$ to 
$10^{-5}\,{\rm M}_\odot/{\rm yr}$ \citep[e.g.,][]{Mayer13,Bujarrabal10}.

The general trend of decreasing cloud brightness, and therefore
decreasing density, continues beyond the ZIMPOL field of view,
as can be inferred from the HST WFC3 image (Fig.~\ref{HSTZimpol}a). 
For example, the H$\alpha$ peak surface brightness of the bright, 
bow shaped jet clouds in the NE at a separation of $\approx 10''$  
($r_{\rm RG}\approx 3\cdot 10^{16}$) is about ten times lower 
than for the clouds H$_{\rm SW}$, I$_{\rm SW}$,
J$_{\rm SW}$ in the ZIMPOL field, and the SB of all 
clouds located at even larger distances $>10''$ are
of the order of 100 times lower.  

\paragraph{Ionized mass and jet outflow limits.} 
Summing up the mass of the bright 
H$\alpha$ clouds and the central source listed in Table~\ref{Nedens} 
yields a summed mass of about $M_{\Sigma{\rm cl}}=5\cdot 10^{-7}{\rm M}_\odot$. 
This can be compared with the expected molecular hydrogen 
gas in the red giant wind. If we assume a mass loss rate 
of $\dot{M}_{\rm RG}=10^{-6}\,{\rm M}_\odot$/yr and $v_\infty=10$~km/s 
or 2 AU/yr (10~mas/yr), then a spherical wind contains about 
$2\cdot 10^{-4}{\rm M}_\odot$ of H$_2$ from 10 to 400~AU. 
This is roughly the region covered by the field of view of ZIMPOL. Thus, the bright H$\alpha$ clouds in R Aqr contain only a 
small fraction of $0.1-1$~\% of the circumstellar matter 
within 400 AU if $\dot{M}_{\rm RG}=10^{-6}{\rm M}_\odot$ is adopted.  

The ionized gas in the 14 selected bright emission regions
fills a volume of about $10^5$~AU, which is of the order of 0.03~\%
of the volume $V_{\rm 400 AU}$ within $r_{\rm RG}=400$~AU from the star. 
All this ionized gas is localized in the two NE and SW jet cones 
with an opening angle of about $\theta\approx \pm 15^\circ$. 
These two cones are about $3~\%$ of the spherical
volume $V_{\rm 400 AU}$. Accordingly, only about 1~\% of the
volume in the adopted $\pm 15^\circ$ jet cones are bright clouds, the
rest of the cone is filled with low emissivity gas. 

We can estimate an upper limit for the low emissivity gas 
which could be present in the two jet cones. The
sensitivity of the ZIMPOL images is quite limited because
of the extended AO-PSF. Ionized gas with the same temperature but 
with ten times lower density (100 times lower emissivity) than
the considered 14 bright H$\alpha$ jet clouds could fill  
the two jet cones without being detected in the ZIMPOL image.
But, the sensitive 
H$\alpha$ and [O~III] HST images (Fig.~\ref{HSTfilters}) 
show faint emission along the two jet cones and this could
originate from such low density gas. 
This ionized low density gas ($T_e\approx 15\,000~$K) could contain 
up to ten times more mass than the bright H$\alpha$ clouds or up to 
$M_{{\rm H}\alpha}\lapprox 10^{-5}\,{\rm M}_\odot$ within $V_{\rm 400 AU}$.
Thus, the observed clouds are perhaps only the higher density 
regions in two ``fully'' ionized jet outflow cones. 

This sets also a limit on the possible mass outflow from the
R Aqr jet. Assuming that the jet outflow consists predominantly
of ionized gas and has a typical outflow velocity of 100~km/s
yields a jet mass loss rate of about 
$\dot{M}_{\rm jet}\lapprox 10^{-7}\,{\rm M}_\odot/{\rm yr}$.  

Unfortunately, only the mass for the bright clouds 
$M_{\Sigma{\rm cl}}=5\cdot 10^{-7}{\rm M}_\odot$ is well measured with
the H$\alpha$ observations.  
All other estimates for the gas mass $M_{{\rm H}_2}$ or $M_{{\rm H}^+}$, 
and mass loss rates $\dot{M}_{\rm RG}$ or $\dot{M}_{\rm jet}$ are very 
uncertain because they depend on observationally not well established
parameters. 
  
\subsection{Other parameters for the jet clouds}

Table~\ref{Nedens} gives for the bright jet clouds
diameters $\oslash_{\rm cl}$, emissivities $\epsilon({\rm H}\alpha)$,
densities $N_e$ , and mass $M_{\rm cl}$ from which we can derive
other interesting cloud properties, like 
H$\alpha$ luminosities $L({\rm H}\alpha)/L_\odot$, 
recombination time scales $t_{\rm rec}$, recombination rates $n_{\rm rec}$, 
and photoionization parameters. We investigate whether besides $N_e$ and
${\rm SB}$ also other cloud properties change systematically 
with distance from the central binary, and compare 
cloud parameters for the central jet source and typical values for 
the five ``inner clouds'' $A_{\rm N}$, $A_{\rm SW}$, $B_{\rm SW}$, 
$C_{\rm SW}$,$D_{\rm SW}$, and the eight ``outer clouds'' 
$B_{\rm NE}$, $C_{\rm NE}$, and $E_{\rm SW}$, $F_{\rm SW}$,
$G_{\rm SW}$, $H_{\rm SW}$, $I_{\rm SW}$, $J_{\rm SW}$. 

Table~\ref{Cloudpara} lists the mean values $m_p={\rm mean}({\rm log} p_i)$
and standard deviations $\sigma_p={\rm stdev}({\rm log} p_i)$
of the logarithmic values for the cloud properties $p_i$ of the two
cloud groups (clouds $i=1,..,n$), and the difference 
$\Delta_p=m_p^{\rm outer}-m_p^{\rm inner}$ between the mean values.  
The logarithmic difference $\Delta_p$ is the ratio between
``outer'' and ``inner'' clouds for a given parameter and it
is equivalent to a power law index for the radial dependence if
divided by $\Delta_r=0.74$ for the cloud distance. For
example the power law index for the density fall-off as derived
in Fig.~\ref{HaRNe} follows from 
$\Delta_{N_e}/\Delta_{r}=-1.35$. The logarithmic difference
reflects also the functional relationship between parameters. For example
for the cloud mass $M_{\rm cl}\propto N_e\cdot V_{\rm cl}$ there is 
$\Delta_{M}=\Delta_{N_e}+\Delta_{V}$. A clear trend for
a cloud parameter is present if 
$|\Delta_p|\gapprox \sigma_p$, while the radial dependence is
weak or not significant for 
$|\Delta_p|\lapprox\sigma_p$. 

Table~\ref{Cloudpara}
gives first the values for the initially measured
cloud parameters $\oslash_{\rm cl}$, SB$({\rm H}\alpha)$,
$F({\rm H}\alpha)$ and then all the deduced parameters. 
A strong trend is present for the surface brightness 
$\Delta_{\rm SB}=1.79$ ($\sigma_{\rm SB}\approx 0.25$) or the cloud
flux, while the trend is only marginal for the cloud diameters 
$\Delta{\rm \oslash}=0.25$ ($\sigma_{\rm \oslash}\approx 0.2$). 
The derived emissivity reflects mainly the strong SB  
dependence $\Delta_\epsilon\approx \Delta_{\rm SB}-\Delta_\oslash$ and
the same applies for the $N_e$-dependence 
$\Delta_{N_e}\approx 0.5\,\Delta_\epsilon$. 


\paragraph{Cloud masses.} 
For the cloud volume, there is a weak trend of cloud sizes with 
separation $\Delta_V\approx 0.8$, but
the scatter $\sigma_V\approx 0.6$ is large. 
There are larger and smaller clouds at all distances 
and the resulting $\Delta_V$-value depends on whether the considered 
sample contains many or only a few small clouds. Thus, the statistics of the 
average cloud volume suffer from a selection effect and
therefore the correlation of $V_{\rm cl}$ with 
distance is uncertain. 

Interesting are the typical cloud masses $M_{\rm cl}$. There exists
a quite large spread of cloud masses in the two groups, but the mean values 
are the same for the ``inner'' and the ``outer'' jet clouds. Cloud masses are
also given in Table~\ref{Nedens} and they are all in the range
$M_{\rm cl}\approx 10^{-9}-10^{-7}~{\rm M}_\odot$. Of course, the sample of measured jet
clouds considers all the bright, more massive clouds and does not
include the faint clouds for which $M_{\rm cl}$ could be 
$<10^{-9}{\rm M}_\odot$. Thus, the correct statement is that the
brightest emission clouds in the ``inner'' jet region and the ``outer'' 
jet region have roughly the same mass.     

\begin{table}
\caption{Logarithmic parameters for the jet source (center), 
and logarithmic mean $m$ and deviation $\sigma$ for the ``inner''
and the ``outer'' H$\alpha$ jet clouds and 
difference $\Delta=m_{\rm outer}-m_{\rm inner}$.}
\label{Cloudpara}
\begin{tabular}{lcccc}
\noalign{\smallskip\hrule\smallskip}
parameter          & center     & ``inner''   & ``outer'' & $\Delta$ \\
\noalign{\smallskip\hrule\smallskip}
log $(r_{\rm RG}/{\rm AU})$  &  1.0  & $1.8\pm 0.2$ & $2.5\pm 0.1$ & +0.74   \\

\noalign{\smallskip measured parameters: diameter, surface brightness, flux \smallskip} 
log $\oslash_{\rm cl}$ [AU]  & 0.8   & $1.2\pm 0.2$ & $1.4\pm 0.2$ & +0.25  \\
log $F_{\rm cl}\tablefootmark{a}$ 
                    & 1.4 & $0.4\pm 0.5$ & $-0.7\pm 0.6$ & $-1.10$ \\
log ${\rm SB}\tablefootmark{b}$   & 4.7 & $2.9\pm 0.2$ & $1.1\pm 0.3$ & $-1.79$ \\

\noalign{\smallskip emissivity, luminosity \smallskip} 
log $\epsilon({\rm H}\alpha)\tablefootmark{c}$  
                    & 2.5  & $ 0.4\pm 0.3$ & $-1.7\pm 0.3$ & $-2.02$ \\
log $(L({\rm H}\alpha)/L_\odot)$ 
                   & $-1.4$ & $-2.5\pm 0.5$ & $-3.7\pm 0.7$ & $-1.28$ \\

\noalign{\smallskip density, volume, mass \smallskip} 
log $N_e$ [cm$^{-3}$] &  7.7  & $6.6\pm 0.1$ & $5.5\pm 0.2$ & $-1.04$   \\
log $(V_{\rm cl}/{\rm AU}^3)$ 
                   &  2.2   & $3.2\pm 0.6$ & $4.0\pm 0.6$ & +0.74   \\   
log $(M_{\rm cl}/M_\odot)$ 
                   & $-7.6$ & $-7.6\pm 0.6$ & $-7.9\pm 0.6$ & $-0.30$ \\
\noalign{\smallskip recombination \smallskip} 
log $(\tau_{\rm rec}/{\rm days})$        
                   & 0.3    & $1.3\pm 0.2$ & $2.3\pm 0.2$ & +0.96 \\
log $n_{\rm rec}$ [s$^{-1}$]
                  & 44.1    & $43.1\pm 0.5$ & $41.8\pm 0.8$ & $-1.26$ \\
\noalign{\smallskip photo-ionization \smallskip} 
log ($\Omega_{\rm cl}/4\pi$)  
                  & 0        & $-1.7\pm 0.5$ & $-2.8\pm 0.5$ & $-1.06$ \\ 
log $Q(H^0)$ [s$^{-1}$]     
                  & 44.1   & $44.8\pm 0.5$ & $44.6\pm 0.3$ & $-0.20$ \\

\noalign{\smallskip\hrule\smallskip}
\end{tabular}
\tablefoottext{a}{$F_{\rm cl}$ in $10^{-12}{\rm erg}\,{\rm cm}^{-2}{\rm s}^{-1}$,} 
\tablefoottext{b}{${\rm SB}_{\rm cl}$ in 
$10^{-12}{\rm erg}\,{\rm cm}^{-2}{\rm s}^{-1}{\rm arcsec}^{-2}$,}
\tablefoottext{c}
{$\epsilon({\rm H}\alpha)$ in $10^{-12}{\rm erg}\,{\rm cm}^{-3}{\rm s}^{-1}$.} 
\end{table}

\paragraph{Ionization parameters.} The H$\alpha$ emission is a good
tracer of the ionized gas. The cloud luminosity $L({\rm H}\alpha)$
is $\propto N_e^2\cdot V_{\rm cl}$ and accordingly $\Delta_L \approx
2\cdot\Delta_{N_e}+\Delta_{V_{\rm cl}}$. The total H$^+$-recombination rates 
per cloud $n_{\rm rec}$ are just proportional to 
the derived H$\alpha$ luminosity $L({\rm H}\alpha)$ of the cloud.

The recombination timescale $\tau_{\rm rec}$ for ionized hydrogen gas 
is for a given $T_e$ inversely proportional to the density 
$\tau_{\rm rec}\propto 1/N_e$. This yields, using the H\,I 
Case$_{\rm B}$-recombination coefficients \citep{Osterbrock06}, very short 
timescales of $\tau_{\rm rec}\approx$ 6 six months for the ``outer'' clouds,
$\approx$ few weeks for the ``inner'' clouds, and only
$\approx$ a few days for the central source. Therefore, H$\alpha$ line
fluxes may be variable on such short timescales. 

The gas ionization of the H$\alpha$-clouds could be caused 
by three processes: (i) photo-ionization by UV-radiation 
from the jet source, (ii) photo-ionization by high energy 
radiation produced by shocks in the jet, and (iii) collisional
ionization in shocked gas that is cooling down in the
post-shock region. It is very possible that all three
processes are involved. A major difference between these
processes is that shock related ionization (ii) and (iii) are
a local effect and each cloud may show an individual temporal 
behavior. Contrary to this the photo-ionization by the central jet source 
is a more global effect and, if dominant, then 
ionization parameters and flux variations
should be reflected in all clouds simultaneously.     

If we assume that the clouds are photo-ionized by the radiation
from the central jet source, then we can estimate the required 
number of ionizing photons to be emitted by the central 
jet source, in order to keep the ionization of the clouds 
in a steady state. Essentially, the recombination rate $n_{\rm rec}$
of a cloud, which is proportional to the cloud luminosity
$L({\rm H}\alpha)$, must be equal to the incoming 
ionizing photons from the jet source. From the cloud distance $r_{\rm jet}$
and diameter $\oslash_{\rm cl}$ follows the angular cross section 
for each cloud $\Omega_{\rm cl}/4\pi$ as seen by the jet source 
and this yields the required ionizing photon luminosity 
$Q({\rm H}^0)=n_{\rm rec} \cdot \Omega_{\rm cl}/4\pi$ for a spherically
emitting jet source keeping the cloud photo-ionized. 

Table~\ref{Cloudpara} gives in the bottom line the $Q({\rm H}^0)$ 
parameter for the jet source, which is almost equal for
the ``inner'' and ``outer'' clouds and just a bit less for the jet 
source itself. This indicates that a source with a spherical 
ionizing photon emission rate of $Q({\rm H}^0)\approx 10^{45}{\rm s}^{-1}$ 
could keep all observed H$\alpha$ clouds within 400~AU ionized. 
For example, a hot subdwarf with a blackbody spectrum 
($T_{\rm eff}=35000~$K, $R=0.1\,{\rm R}_\odot$, 
$L= 13\,{\rm L}_\odot$) produces log~$Q\approx 45$ ionizing photons/s. If the ionizing photons are only emitted inside the two
jet cones with an opening angle of $\theta_{\rm cone}\pm 15^\circ$, then the
number of required ionizing photons is reduced by a factor 
$F_{\rm cone}/F_{4\pi}=1-{\rm cos}\theta_{\rm cone}=0.034$
to $Q_{\rm cone}({\rm H}^0)\approx 10^{43.5}{\rm s}^{-1}$. 

The fact that the determined minimum $Q({\rm H}^0)$ is the same 
for the ``inner'' and ``outer'' jet clouds and similar for the
central jet source suggests strongly that photo-ionization by
the hot stellar source is important. If true, then brightness
changes of the accreting hot source should induce H$\alpha$ brightness 
variations for a whole group of clouds at the same time. 

Another aspect of photo-ionization by the central jet source is the
shadowing caused by the red giant companion and its wind. 
This might explain why the H$\alpha$-clouds in the NE observed 
in 1991 by \citep{Paresce94} had on average a larger position angle 
than the clouds observed in 2014
(see Sect.~\ref{Sectmotion}). NE-clouds located at  
$\theta>30^\circ$ are perhaps absent in our data because 
the ionizing radiation from the 
jet source cannot reach such clouds for the current configuration
of the binary system. 

The ionizing photons from an accreting jet source could be emitted 
into two opposite directions producing two ionization 
cones as proposed by \citet{Kafatos86}. The geometric appearence 
of the innermost $r_{\rm jet}<400~$AU, possibly photonionized, 
H$\alpha$-clouds could therefore be strongly influenced by this 
collimated irradiation effect and we are seeing just
the dense gas in this ionization cone irrespective of whether
it is gas in the slow red giant wind or gas in the fast jet outflow.  
Observing the cloud motion with repeated observations will clarify
this important issue. 

\subsection{Comparison with previous parameter determinations}

Physical parameters for the emission nebula in R Aqr have been 
determined in many previous studies. These determinations are 
mainly based on emission line spectroscopy
in the visual range with ground based observations and in 
the UV with the International Ultraviolet Explorer (IUE) satellite 
\citep[e.g.,][]{Wallerstein80,Kaler81,
Michalitsianos82,Kafatos86,Hollis91,Burgarella92,Meier95}. These spectra 
cover the central region of R Aqr or/and the NE and SW jet 
features at separations of $\approx 5''$
with typically a spatial resolution of a few arcsec. 

The different studies obtained quite consistent nebular parameters of
roughly $N_e\approx 10^6{\rm cm}^{-3}$ and $T_e\approx 20000~$K
for the central H~II region based on different types of diagnostic
measurements, mainly emission line ratios. These parameters are
quite similar to a kind of average value for the jet clouds
at $r_{\rm RG}<400~$AU ($<2''$) derived in this work. 

Also in good agreement are the literature values for the
NE and SW jets at separations of $\approx 1000$~AU with
derived densities of about $N_e\approx 10^4{\rm cm}^{-3}$ for the
NE jet and $N_e\approx 10^3{\rm cm}^{-3}$ for the SE jet and 
there are clear signs for the presence of high temperature gas $T_e>20000$~K, 
which emits O~VI lines and strongly variable 
X-rays emission \citep{Kellogg07,Nichols09}. 
The NE and SW jet emissions at $\gapprox 1000$~AU are explained  
as radiative shocks because of the collision of fast jet gas with
slow gas clouds.  

The radiation parameters of the central jet source were 
estimated with, for example, the Zanstra method, based on the assumption 
that the central emission nebula is predominantly photo-ionized by 
this source. Suggested ionizing sources are either a hot white dwarf 
$T\gapprox 50\,000$~K \citep[e.g.,][]{Meier95}, or a hot subdwarf
$T\approx 40\,000~K$, $L\approx 10~{\rm L}_\odot$, 
$R\approx 0.1\,{\rm R}_\odot$ \citep{Burgarella92}. In
alternative models the UV radiation is proposed to originate  
from the accretion disk and the boundary layer 
around a white dwarf or subdwarf with an accretion rate of the order 
$\dot{M}_{\rm acc}\approx 10^{-8}\,{\rm M}_\odot{\rm yr}^{-1}$ 
\citep[e.g.,][]{Burgarella92}. Estimated values for the number 
of ionizing photons emitted by the central source are 
log~$Q\approx 42.5$ \citep{Meier95}.

\section{Summary and outlook}

This work describes and analyzes the innermost region of the 
\object{R Aqr} jet based on
new line filter images from SPHERE-ZIMPOL and HST - WFC3. The most
important results from these data are the resolution of the R Aqr binary
system, a quantitative analysis of the H$\alpha$ emission line clouds, and
new insights on the structure and the physics of the jet outflow. These
main topics are addressed in the following discussion.  

\subsection{Binary orientation and orbit}

The presented SPHERE-ZIMPOL images taken in the H$\alpha$ filters resolve 
the jet source and the mira variable in the R Aqr binary for 
the first time unambiguously and with high astrometric precision. 
It can be expected that successful re-observations of the stellar binary
are easy to achieve near photometric minimum allowing for a 
fast progress in the determination of an accurate orbit in the future. 

We know up to now only an approximate orbital period
of about 44 years from three obscuration events \citep{Willson81}
and a not well defined, because hard to measure, radial velocity
curve for the mira variable confirming the 44 years periodicity
\citep[e.g.,][]{Gromadzki09a}. A better knowledge of the orbit and 
especially of the orientation of the orbital plane is essential for 
an interpretation of the geometry of the jet and the extended nebulosity, 
and the determination of stellar masses.

The R Aqr system was probably resolved previously by \citet{Hollis97b}
with high resolution radio interferometry taken in November 1996. 
They could measure the position of the peak of a slightly 
extended ($\approx 100$~mas) 7 mm continuum emission located 
about 55~mas N-NE ($18^\circ$)
from a point-like (diameter $<30$ mas) SiO maser emission. 
These two emission components were associated with the two stars 
in the binary, the continuum emission with the ionized gas surrounding the 
accreting jet source and the SiO emission with the dense 
envelope of the mira variable. But, there are some doubts
as to whether the peak of the extended continuum emission from 1996 
represents the jet source, because the
derived separation between the two stars is surprisingly large 
for the 1996 epoch. This would imply a large orbital
eccentricity and relatively large stellar masses \citep{Hollis97b},
which are not in agreement with the later derived
radial velocity curve for the red giant by \citet{Gromadzki09a}.
\citet{Ragland08} suspected that the measured radio continuum peak may
not coincide well with the stellar source, and the H$\alpha$ maps presented
in this work show bright jet clouds near the jet source. Such
a cloud might have been responsible for the peak emission seen
by \citet{Hollis97b}. However, we may still assume that the jet 
source and innermost jet clouds were located
roughly north of the red giant in 1996. If correct, then the 
stellar positions measured now indicate a binary orbit in clockwise 
direction, from north over to west.

The orientation of the binary is also important for the interpretation
of high resolution maser line maps of
R Aqr \citep[e.g.,][]{Boboltz97,Hollis00,Cotton06,Ragland08,Kamohara10}.
The existence of a symmetry axis has been suggested repeatedly for the
distribution of the maser spots which could be related to the rotation axis
of the mira. 
If the hot companion has an impact on the 
maser emission in the envelope of the mira then a strong east - west
asymmetry could be apparent for the current binary orientation.  

The radial velocity curve of \citet{Gromadzki09a} predicts 
that the hot accreting component will pass in the coming ten years 
in front of the mira variable with a relative angular velocity 
of about 10 mas/year. This relative motions should be easily 
measurable with SPHERE-ZIMPOL and provide, in connection with
high precision astrometry of the mira from 
radio interferometry \citep[e.g.,][]{Min14} or the GAIA satellite,
accurate estimates on the masses of the two stellar components. 
 
\subsection{Flux measurements for the H$\alpha$ clouds}

Absolute flux measurements for complex sources, like the H$\alpha$
jet clouds around the strongly variable object R Aqr, 
are difficult to achieve with ground based, extreme adaptive 
optics measurements 
because of the strong PSF variations caused by the rapidly
changing atmospheric conditions and AO performance. 
The unique coincidence of quasi-simultaneous HST data provides
for our ZIMPOL - SPHERE test data an ideal opportunity for assessing 
and checking the quality of our measuring procedure and instrument
calibrations.

The ZIMPOL H$\alpha$ cloud flux measurements require an analysis 
of the PSF to account for the non-focussed flux in the extended PSF halo. 
The derived correction factors are large, of the order of three to ten
typically, and depend on the cloud and aperture size. This is the 
major source of uncertainty in the absolute flux measurements. The resulting 
line fluxes from ZIMPOL - SPHERE are about 30~\% higher than the
fluxes derived from the HST data. The main reason for this
discrepancy is most likely a systematic
overestimation of the flux correction factors for the ZIMPOL data. 

Flux ratios between different clouds agree very well
$\sigma \approx 10~\%$ between 
different ZIMPOL observing dates and different 
ZIMPOL H$\alpha$ filters. The agreement
is also excellent $\sigma = 7.5~\%$ between relative cloud fluxes measured 
in ZIMPOL data and HST data. This indicates that line flux variation can
be measured with high sensitivity with ZIMPOL - SPHERE, if one emission
component in the field can be used as flux reference.  

The absolute H$\alpha$ line flux is a very important diagnostic
for the ionized gas in spatially resolved clouds. The H$\alpha$ fluxes 
$F({\rm H}\alpha)$ provides, together with the 
diameters $\oslash_{\rm cl}$ of the  
emitting cloud, the average emissivities 
$\langle \epsilon({\rm H}\alpha)\rangle$
which scales like $\epsilon({\rm H}\alpha)\propto N_e^2$ with the 
density (with only a weak $T_e$-dependence) and, together with
$\oslash_{\rm cl}$ , we can also derive cloud masses $M_{\rm cl}$, 
recombination rates, recombination timescales, 
and estimates on the cloud energy budgets. 
Our procedure to determine the cloud densities in the jet of 
R Aqr from $\epsilon({\rm H}\alpha)$ is
confirmed by the HST [O~III] line observations, which show for
high density clouds $N_e>10^6{\rm cm}^{-3}$ the expected low [O~III]/H$\alpha$ 
line ratio because of the collisional deexcitation. 
 
\subsection{On the R Aqr jet}

The presented SPHERE-ZIMPOL data and WFC3 - HST data provide 
two most significant advances for the investigation
of the R Aqr jet with respect to previous HST imaging
observations \citep{Paresce91,Paresce94,Hollis97a}.
First, the SPHERE-ZIMPOL
images provide a spatial resolution of about 25~mas (FWHM), 
which is about three times higher than for the previous and 
new HST observations. Therefore the SPHERE-ZIMPOL data can
resolve the central binary and the innermost jet clouds. 
Second, the simultaneous WFC3 data provide very sensitive, wide field 
images of the R Aqr jet for four important diagnostic lines. 
This combination of high
quality data forms a very rich source of observational information
about the physical properties of the R Aqr jet. In this work,
we focus the analysis mainly on the H$\alpha$ emission of the
innermost jet $r_{\rm jet} < 400$~AU covered by the ZIMPOL data. 
 
A very important aspect of the R Aqr jet is the proximity of this
system allowing us to resolve jet cloud structures on scales of
a $\sim 5$~AU and measure position with a relative precision of
$\sim 1$~AU. We detect a large diversity of cloud structures, 
like point-like unresolved clouds, elliptical and
bubble-like features, and short straight (transverse) and long
undulating (radial) filaments. In addition the jet clouds
are very bright in H$\alpha$ and it is possible to
see a large range of bright and faint clouds in the outflow.
All these features are expected to move and evolve in brightness
and shape within a few years. R Aqr is for these reasons 
a unique laboratory for the investigation of hydrodynamical 
processes of jet outflows.

Jets are a frequent phenomenon in symbiotic binaries. Other
well known examples are the precessing jets in \object{CH Cyg}, 
which resemble in
many respects the R Aqr jets with radio emissions 
\citep[e.g.,][]{Taylor86,Crocker02}, 
H$\alpha$ and [O~III] line clouds, and X-ray hot spots 
\citep{Corradi01,Karovska10}. Other spectacular examples are
\object{Sanduleak's star} in the Large Magellanic Clouds with a jet 
that extends over 14~pc \citep{Angeloni11}, or \object{MWC~560}, where
the line of sight is parallel to the jet axis and the outflowing
gas is seen in absorption in the continuum of the jet source 
\citep{Tomov90,Schmid01}. Because of the orbital motion and the interaction with
the companion, the jets in symbiotic binaries are expected
to precess or wobble. This could also explain some
of the structures seen in R Aqr.
The R Aqr jet might also provide complementary information
with respect to imaging studies of famous jet outflows from young
stars like HH1/2, HH34, HH47, and others 
\citep[see, e.g.,][]{Raga16,Hartigan11}.

For the R Aqr jet we have derived from the H$\alpha$ emission 
a very high gas density $N_e$ with a clear radial dependence 
from $N_e\approx 5\cdot 10^7\,{\rm cm}^{-3}$ for the central jet source
to $N_e\approx 4\cdot 10^6\,{\rm cm}^{-3}$ for the ``inner'' jet clouds
at a separation of about 60~AU, to $N_e\approx 3\cdot 10^5{\rm cm}^{-3}$  
for the ``outer'' clouds around 300~AU. 
Because of the high density, the recombination
timescales are very short for the jet clouds, less than a
week for the central source, about three weeks for the ``inner'' 
clouds, and about six months for the ``outer'' clouds, indicating
that one should expect flux variations on such timescales. 

Other cloud parameters with a strong
anti-correlation with cloud distance are the H$\alpha$ flux 
surface brightness, or luminosity. The masses of the
clouds show no clear trend, while the cloud diameters have a tendency
to increase with distance.  

We should expect that the jet in R Aqr is very dynamic. 
Because of the orbital motion the fast outflow from the 
jet source must move through the dense
stellar wind causing certainly complex hydrodynamical
processes which evolve on timescales of about a year or shorter. 
Therefore, it is not surprising that the cloud geometry 
has completely changed within $r_{\rm jet}<400$ AU when compared 
to the HST observations from 1991 and 1992 presented by \citet{Paresce94}. 
Further out, at $r_{\rm jet}\approx 1000-2000$~AU, 
the jet clouds are less variable and the
evolution and motion of certain features can be followed
over decades \citep{Navarro03}.

A most important task for future observations is the 
determination of the orientation of the orbital plane. 
This will help us to understand the interaction between
the jet originating from an orbiting source with the
red giant wind and answer the important question
whether the jet outflow is oriented perpendicular to the orbital 
plane or not. From an inclined outflow one may expect 
jet precession that could explain the partly
point-symmetric arc structures associated with the NE-SW jet directions. 

The data presented in this work provide a wealth of
new information for the detailed investigation of the
physical nature of the jet clouds in the innermost region of
the R Aqr jet. For example, it would be interesting to know whether
the H$\alpha$ clouds are ionized gas regions
of the slow red giant wind, which are excited by shocks from the 
fast outflow (e.g., the clouds in the SW). Alternatively, 
they might also be radiative shock
regions moving with the fast outflow (e.g., the tangential wisps 
B$_{\rm NE}$ and C$_{\rm NE}$), or dense clouds from
the dense wind which are entrained and accelerated 
outwards in the fast jet outflow (perhaps the long filament A$_N$).

In this work we have derived many cloud parameters for 
more than a dozen jet features. It is beyond the scope of this 
first observational paper to investigate the nature of the 
individual clouds in more detail. Very important
additional information about the individual clouds 
can be gained from future observations which should reveal 
the motion and flux variations of individual clouds. Such a
cloud monitoring may provide the key for the
interpretation of the observed diverse individual 
cloud structures in terms of jet hydrodynamics and shock
physics. We plan to carry out such re-observations of
the R Aqr system with the goal of advancing significantly
our understanding of the physics of stellar jet outflows.

\begin{acknowledgements}
SPHERE is an instrument designed and built by a consortium consisting
of IPAG (Grenoble, France), MPIA (Heidelberg, Germany), LAM (Marseille,
France), LESIA (Paris, France), Laboratoire Lagrange (Nice, France), 
INAF – Osservatorio di Padova (Italy), Observatoire de Gen`eve 
(Switzerland), ETH Zurich (Switzerland), NOVA (Netherlands), ONERA (France), 
and ASTRON (Netherlands), in collaboration with ESO. 
SPHERE was funded by ESO, with additional contributions from CNRS (France), 
MPIA (Germany), INAF (Italy), FINES (Switzerland), and NOVA (Netherlands). 
SPHERE also received funding from the European Commission Sixth and 
Seventh Framework Programmes as part of the Optical Infrared 
Coordination Network for Astronomy (OPTICON)
under grant number RII3-Ct-2004-001566 for FP6 (2004–2008), grant number
226604 for FP7 (2009–2012), and grant number 312430 for FP7 (2013–2016).

This work has been carried out within the frame of the National 
Centre for Competence in Research ``PlanetS'' supported by the Swiss 
National Science Foundation. H.M.S., S.Q., and C.T. acknowledge 
the financial support of the SNSF.

A.B., R.C, S.D., R.G., B.S., E.S., and M.T. acknowledge support from the
``Progetti Premiali'' funding scheme of the Italian Ministry of Education,
University and Research. 

This research has made use of the SIMBAD database, operated
at CDS, Strasbourg, France. 
\end{acknowledgements}

\bibliographystyle{aa} 
\bibliography{RAqr_arXiv.bib} 

\end{document}